\documentclass[aip]{revtex4-1}

\usepackage{graphicx}
\usepackage{amsmath,amssymb,physics,dsfont}
\usepackage{color}
\usepackage{hyperref}
\usepackage{microtype}
\usepackage{braket}
\usepackage{defs-private}
\usepackage{qcircuit}
\usepackage{tikz}

\usepackage[utf8]{inputenc}
\usepackage[T1]{fontenc}
\usepackage{mathptmx}
\usepackage{etoolbox}

\makeatletter
\def\@email#1#2{%
 \endgroup
 \patchcmd{\titleblock@produce}
  {\frontmatter@RRAPformat}
  {\frontmatter@RRAPformat{\produce@RRAP{*#1\href{mailto:#2}{#2}}}\frontmatter@RRAPformat}
  {}{}
}%

\begin{document}

\title{Efficient Berry Phase Calculation via Adaptive Variational Quantum Computing Approach}

\author{Martin Mootz}
\email{mootz@iastate.edu}
\affiliation{Ames National Laboratory, U.S. Department of Energy, Ames, Iowa 50011, USA}

\author{Yong-Xin Yao}
\email{ykent@iastate.edu}
\affiliation{Ames National Laboratory, U.S. Department of Energy, Ames, Iowa 50011, USA}
\affiliation{Department of Physics and Astronomy, Iowa State University, Ames, Iowa 50011, USA}

\begin{abstract}
We present an adaptive variational quantum algorithm to estimate the Berry phase accumulated by a nondegenerate ground state under cyclic, adiabatic evolution of a time-dependent Hamiltonian. Our method leverages cyclic adiabatic evolution of the Hamiltonian and employs adaptive variational quantum algorithms for state preparation and evolution, optimizing circuit efficiency while maintaining high accuracy. We benchmark our approach on dimerized Fermi-Hubbard chains with four sites, demonstrating precise Berry phase simulations in both noninteracting and interacting regimes. Our results show that circuit depths reach up to 106 layers for noninteracting systems and increase to 279 layers for interacting systems due to added complexity. Additionally, we demonstrate the robustness of our scheme across a wide range of parameters governing adiabatic evolution and variational algorithm. These findings highlight the potential of adaptive variational quantum algorithms for advancing quantum simulations of topological materials and computing geometric phases in strongly correlated systems.
\end{abstract}

\maketitle

\section{Introduction}

The Berry phase is a fundamental geometric property of quantum states and plays a crucial role in characterizing topological phases of matter~\cite{Berry1984,Xiao2010}. Since its introduction, it has been recognized as a key quantity underlying diverse physical phenomena, including the quantum Hall effect~\cite{Thouless1982,Niu1985}, topological insulators~\cite{Hasan2010,Qi2011}, and superconductors~\cite{Qi2011,Ando2015}. Computing the Berry phase and related topological invariants, such as the Chern number~\cite{Armitage2018} and Zak phase~\cite{Zak1989,Resta1994}, is essential for distinguishing different topological phases in condensed matter. However, accurately computing these quantities, particularly in strongly correlated and large-scale systems, poses a significant challenge for classical computational methods.
Traditional approaches to computing the Berry phase for many-body systems often rely on numerical diagonalization~\cite{Resta2000,Marzari2012}, Monte Carlo simulations~\cite{Motoyama2013,Kolodrubetz2014,Yamamoto2016}, and tensor network techniques~\cite{Pollmann2012,Bultinck2018,Ohyama2025}, which are often limited in practical applications due to exponential scaling of computational load, fermionic  sign problems in sampling, or entanglement growth necessitating prohibitively high bond dimension or contraction costs.

Quantum computing provides a promising route for efficiently computing the Berry phase and other topological invariants, as it inherently captures quantum correlations and evolution~\cite{feynman82qc,asp_ipea,kimEvidenceUtilityQuantum2023}. Recent advances in noisy intermediate-scale quantum (NISQ) devices have enabled the development of quantum algorithms for probing topological properties of matter~\cite{nisq,quantum_supremacy19}. Several studies have demonstrated the feasibility of using quantum circuits to identify topological phases~\cite{Roushan2014,Zhan2017,Choo2018,Azses2020,Mei2020,Xiao2021determiningquantum,Smith2022,Koh2024}, distinguish different topological orders~\cite{KITAEV20032,Wen2017}, and compute topological invariants~\cite{Flurin2017,Xu2018,Elben2020,Niedermeier2024,sugimoto2024}. Although current quantum hardware is limited by noise and decoherence, these works highlight the potential of quantum computing algorithms to overcome the scalability challenges faced by classical methods.

Various quantum computing approaches have been proposed for Berry phase calculations. Quantum quench dynamics approaches~\cite{Chang2022,Chang2025} extract topological information by analyzing real-time evolution following sudden parameter changes. Variational quantum eigensolver (VQE)-based methods~\cite{Tamiya2021,Ciaramelletti2025} have also been explored for characterizing wave function topology and have been successfully applied to calculating topological invariants, including the Berry phase, on NISQ devices~\cite{Xiao2023}. Alternatively, cyclic adiabatic evolution~\cite{Murta2020,Niedermeier2024} measures the Berry phase by tracking the phase accumulated by a quantum system as it adiabatically follows an instantaneous eigenstate while Hamiltonian parameters evolve slowly along a closed path. This method has been implemented in both single-particle and interacting models using a Trotter approach for adiabatic evolution~\cite{Murta2020}. However, Trotterization of adiabatic state evolution can generate deep quantum circuits, making it challenging to implement on near-term quantum devices.

To alleviate the circuit depth constraints, the adaptive variational quantum dynamics simulation (AVQDS) algorithm provides a useful alternative for simulating adiabatic evolution~\cite{AVQDS}. Unlike fixed-ansatz variational methods, AVQDS dynamically constructs an efficient variational ansatz by selecting parameterized unitaries from a predefined operator pool. Specifically, during the time evolution, unitaries are added to the variational ansatz to keep the McLachlan distance~\cite{mclachlan64variational}, which quantifies deviations between the variational and exact state evolution, below a predefined threshold. By leveraging this strategy, AVQDS reduces quantum circuit complexity while maintaining simulation accuracy, making it well-suited for quantum simulations of topological properties.

In this work, we extend the Trotterized cyclic adiabatic evolution approach in Ref.~\citenum{Murta2020} with AVQDS, demonstrating its efficiency in computing the Berry phase for both noninteracting and interacting systems. To benchmark our method, we compute the Berry phase for a dimerized Fermi-Hubbard chain with four sites and compare our results to exact diagonalization. We analyze quantum circuit complexity and demonstrate the robustness of our approach across a broad parameter space pertaining to adiabatic evolution and variational algorithm. Our findings underscore the potential of AVQDS for efficient quantum simulations of topological materials.

The paper is organized as follows. In Sec.~\ref{sec:avqds}, we provide an overview of the AVQDS algorithm. Sec.~\ref{sec:methods} discusses our methodology for computing the Berry phase, including quantum circuit implementation. Sec.~\ref{sec:res} presents our numerical results for the dimerized Fermi-Hubbard model in the noninteracting and interacting regimes, benchmarking AVQDS against exact diagonalization as well as analyzing computational cost and robustness. Finally, Sec.~\ref{sec:con} presents our conclusions and outlines future directions for extending AVQDS to more complex topological systems.

\section{AVQDS algorithm}
\label{sec:avqds}

In this section, we provide a brief overview of the AVQDS algorithm while a more detailed discussion can be found in our previous works~\cite{AVQDS,mootz2023twodimensional,mootz2024adaptive}. We consider a quantum system described by the pure state $\ket{\Psi}$ which evolves under a time-dependent Hamiltonian $\h$. The system's dynamics is governed by the von Neumann equation~\cite{berman1991Liouville-vonNeumann}, which describes the time evolution of the density matrix $\rho = \ket{\Psi}\bra{\Psi}$:
\begin{align}
	\frac{\mathrm{d}\rho}{\mathrm{d}t}=-\mathrm{i}\left[\hat{\mathcal{H}},\rho\right]\,.
\end{align}
In variational quantum simulations, the quantum state is represented as $\ket{\Psi[\bth]}$, where the real-valued vector $\bth(t)$ comprises $N_\theta$ variational parameters that evolve over time. To determine their dynamics, one employs the McLachlan variational principle~\cite{mclachlan64variational}, which minimizes the squared McLachlan distance. This distance quantifies the discrepancy between the exact and variational time evolution and is expressed in terms of the Frobenius norm ($| \mathcal{O} | \equiv \Tr[\mathcal{O}^\dag \mathcal{O}]$) as
\begin{align}
\label{eq:L2}
    \mathcal{L}^2&\equiv\bigg\|\sum_\mu\frac{\partial\rho[\bth]}{\partial\theta_\mu}\dot{\theta}_\mu+\mathrm{i}\left[\hat{\mathcal{H}},\rho\right]\bigg\|^2 \nonumber \\
    &=\sum_{\mu\nu}M_{\mu\nu}\dot{\theta}_\mu\dot{\theta}_\nu-2\sum_\mu V_\mu\dot{\theta}_\mu+2\,\mathrm{var}_{\bth}[\h]\,.
\end{align}
Here, $M$ is an $N_\theta\times N_\theta$ matrix, and $V$ is a vector of dimension $N_\theta$: 
\begin{align}
\label{eq:M}
M_{\mu,\nu}\equiv &\,\mathrm{Tr}\left[\frac{\partial\rho[\bth]}{\partial\theta_\mu}\frac{\partial\rho[\bth]}{\partial\theta_\nu}\right]=2\,\mathrm{Re}\left[\frac{\partial\langle\Psi[\bth]|}{\partial \theta_\mu}\frac{\partial|\Psi[\bth]\rangle}{\partial \theta_\nu}\right.\nonumber \\ &\left.+\frac{\partial\langle\Psi[\bth]|}{\partial \theta_\mu}|\Psi[\bth]\rangle\frac{\partial\langle\Psi[\bth]|}{\partial \theta_\nu}|\Psi[\bth]\rangle\right]\,, \\
	V_\mu=&\,2\,\mathrm{Im}\left[\frac{\partial\langle\Psi[\bth]|}{\partial \theta_\mu}\h|\Psi[\bth]\rangle+\langle\Psi[\bth]|\frac{\partial|\Psi[\bth]\rangle}{\partial \theta_\mu}\langle \h\rangle_{\bth}\right]\,,
 \label{eq:V}
\end{align}
where $\langle \h\rangle_{\bth}=\langle\Psi[\bth]|\h|\Psi[\bth]\rangle$ and $\mathrm{var}_{\bth}[\h]=\langle\h^2\rangle_{\bth}-\langle\h\rangle^2_{\bth}$ denote the expectation value and variance of $\h$ in the variational state $\ket{\Psi[\bth]}$, respectively. The second terms in both $M$ and $V$ arise from the global phase contribution~\cite{theory_vqs}, with $M$ corresponding to the quantum Fisher information matrix~\cite{meyer2021fisher}. Although our density matrix formulation removes the need to explicitly track the global phase $\varphi_\mathrm{G}$ of $\ket{\Psi[\bth]}$, its determination remains essential for the Berry phase computation using the quantum approach discussed in Sec.~\ref{sec:methods}. The global phase follows the equation of motion~\cite{theory_vqs,Zoufal2023}:
 \begin{align}
 \label{eq:phiG}
     \dot{\varphi_\mathrm{G}}=-\langle \h\rangle_{\bth}+\mathrm{Im}\left[\sum_\mu \langle\Psi[\bth]|\frac{\partial|\Psi[\bth]\rangle}{\partial \theta_\mu}\dot{\theta}_\mu\right]\,,
 \end{align}
where the first term on the right-hand side corresponds to the dynamical phase, while the second term captures the geometric properties of the quantum evolution.
Minimizing $\mathcal{L}^2$ with respect to $\{\dot{\theta}_\mu\}$ yields the equation of motion for the variational parameters:
\begin{align}
\label{eq:theta_eom}
    \sum_\nu M_{\mu\nu}\dot{\theta}_\nu=V_\mu\,.
\end{align}
 The optimized McLachlan distance for the variational ansatz 
$|\Psi[\bth]\rangle$ is given by
\begin{align}
    L^2=2\,\mathrm{var}_{\bth}[\h]-\sum_{\mu}V_\mu \dot{\theta}_\mu\,,
\end{align}
which serves as a measure of the accuracy of the variational dynamics. 

The AVQDS algorithm dynamically constructs a compact variational ansatz in a pseudo-Trotter form: 
\begin{align}
\label{eq:ansatz}
|\Psi[\bth]\rangle=\prod_{\mu=1}^{N_\theta}\mathrm{e}^{-\mathrm{i}\theta_\mu\hat{\mathcal{A}}_\mu}|\Psi_0\rangle\,,
\end{align}
where $\hat{\mathcal{A}}_\mu$ are Hermitian generators and $|\Psi_0\rangle$ is the reference state. Unlike fixed-ansatz variational methods, the AVQDS approach keeps the McLachlan distance $L^2$ below a predefined threshold $L^2_\text{cut}$ during the evolution by dynamically selecting operators $\hat{\mathcal{A}}_\mu$ from a predefined pool and appending them to the ansatz~\eqref{eq:ansatz}. In the original AVQDS approach~\cite{AVQDS}, operators are appended individually in each iteration. In this work, we follow the strategy outlined in Ref.~\citenum{Zhang2025TETRIS} where multiple operators are appended simultaneously, ensuring they act on disjoint sets of qubits. This modification significantly reduces circuit depth while only marginally changing the total number of CNOT gates. 
Specifically, at each iteration, the McLachlan distance $L^2_\nu$ is evaluated for an ansatz of the form $e^{-i\theta_\nu\hat{\mathcal{A}}_\nu}\ket{\Psi[\bth]}$ for all $\hat{\mathcal{A}}_\nu$ in the operator pool of size $N_\mathrm{p}$. Operators are ranked by their ability to reduce $L^2$ and the operator with the smallest $L^2_\nu$ is appended to the ansatz~\eqref{eq:ansatz} as the first addition in the iteration. Subsequently, additional operators are appended with progressively larger or equal $L^2_\nu$, ensuring that each newly appended operator has strictly disjoint support from those selected earlier in the iteration. This process continues until all qubits are covered or no more suitable operators can be found. Importantly, newly appended operators have their variational parameters initialized to zero to ensure smooth wavefunction evolution. This initialization preserves the system’s state but can still reduce the McLachlan distance $L^2$ because of the nonzero derivative with respect to $\theta_\nu$. After updating the ansatz, the McLachlan distance $L^2$ is recalculated and compared against $L^2_\text{cut}$. The adaptive ansatz growth procedure continues until $L^2$ falls below the threshold $L_\mathrm{cut}^2$.

In the simulations, the equations of motion for the variational parameters, Eq.~\eqref{eq:theta_eom}, are integrated using a fourth-order Runge-Kutta (RK4) method, which provides high accuracy while allowing for larger time steps compared to lower-order methods such as Euler integration~\cite{mootz2023twodimensional}. The time step $\delta t$ is adaptively chosen to ensure small parameter updates, specifically requiring $ \max_{\mu}|\dot{\theta}_\mu\delta t|<\delta\theta_\mathrm{max}$, with $\delta\theta_\mathrm{max}=0.01$, a choice that yields accurate quantum dynamics. When not otherwise stated, we use a McLachlan cutoff of $L^2_\text{cut}=10^{-4}$ to ensure close agreement between the variational and exact quantum dynamics. 

For completeness, we briefly outline the circuit-complexity analysis for AVQDS as used here; see Refs.~\citenum{AVQDS,AVQITE} for full complexity derivations of AVQDS and AVQITE. In contrast to the first-order Euler integrator used in Refs.~\citenum{AVQDS,AVQITE}, we adopt the RK4 scheme discussed above. Let $N_\theta$ be the number of variational parameters in the current ansatz and $N_{\mathrm H}$ the number of Pauli strings in the Hamiltonian (which also equals the operator-pool size in this work). Following Ref.~\citenum{AVQDS}, the numbers of distinct direct-measurement and generalized Hadamard-test circuits needed for one evaluation of $\{M, V, \langle \h\rangle_\theta, \langle \h^2\rangle_\theta\}$ are upper bounded by $(N_{\mathrm H}+2)N_\theta + N_{\mathrm H} + N_{\mathrm H}^2$ and $
N_{\mathrm H}(N_\theta-1) + \tfrac{1}{2}N_\theta(N_\theta-1)$, respectively. Defining the per-evaluation total $N_{\mathrm{circ}}^{(\mathrm{eval})} = (N_{\mathrm H}+2)N_\theta + N_{\mathrm H} + N_{\mathrm H}^2 + N_{\mathrm H}(N_\theta-1) + \tfrac{1}{2}N_\theta(N_\theta-1)$, one RK4 step performs four such evaluations and therefore uses $4\,N_{\mathrm{circ}}^{(\mathrm{eval})}$ circuits. If the ansatz expands in a given iteration, an additional $N_{\mathrm H}(N_\theta-1)$ generalized Hadamard-test circuits are incurred. Writing $N_{\text{steps}}$ for the number of RK4 steps along the loop, the overall quantum cost scales as $
O\!\big(N_{\text{steps}}\,N_{\mathrm{circ}}^{(\mathrm{eval})}\big)$,
with $N_{\text{steps}}=T/\delta t$ for fixed step size, or set adaptively by the tolerance $ \max_{\mu}|\dot{\theta}_\mu\delta t|<\delta\theta_\mathrm{max}$.

For the benchmarks presented in Sec.~\ref{sec:res}, we have $N_{\mathrm H}=18$ for the noninteracting Fermi-Hubbard model ($U=0$) and $N_{\mathrm H}=22$ for the interacting Fermi-Hubbard model ($U=10$). Using the observed end-of-loop parameter counts $N_\theta$, for $U=0$ we have $N_\theta=\{74,153,115\}$ at evolution times $T=\{20,100,200\}$. The corresponding per-evaluation circuit upper bounds are $\{5.84\times 10^{3},\,1.78\times 10^{4},\,1.12\times 10^{4}\}$, with additional growth costs $\{1.31\times 10^{3},\,2.74\times 10^{3},\,2.05\times 10^{3}\}$. For $U=10$ with $N_\theta=\{187,184,197\}$, the per-evaluation bounds are $\{2.65\times 10^{4},\,2.58\times 10^{4},\,2.89\times 10^{4}\}$ and the growth costs are $\{4.09\times 10^{3},\,4.03\times 10^{3},\,4.31\times 10^{3}\}$. Each RK4 step uses $4\times$ the per-evaluation count.

To assess the accuracy of AVQDS, we compare its performance against exact diagonalization (ED), where the state propagation follows  
\begin{align} |\Psi[t+\delta t]\rangle = \mathrm{e}^{-\mathrm{i}\delta t \hat{\mathcal{H}}}|\Psi[t]\rangle\,. \label{eq:ED} 
\end{align} 
Note that $\hat{\mathcal{H}}$ is implicitly time-dependent. 
The ED simulations are carried out on a uniform time grid with a step size of $\delta t=0.001$, yielding numerically converged results.

\section{Method \label{sec:methods}}

In this section, we outline our approach for computing the Berry phase in topological systems using cyclic adiabatic evolution. This approach efficiently determines the Berry phase by leveraging adaptive variational methods for state preparation and evolution, thereby avoiding the deep quantum circuits required in standard Trotterization methods.

\subsection{Cyclic adiabatic evolution}

In cyclic adiabatic evolution, a quantum system evolves under a Hamiltonian $\h(\lambda)$ that depends on a periodic, time-dependent parameter $\lambda(t)$, which can be a twist angle for fermionic modes as discussed later. To ensure the validity of this approach,  the Hamiltonian $\h(\lambda)$ must maintain a nondegenerate ground state $\ket{G(\lambda)}$ with energy $E_\mathrm{G}(\lambda)$ throughout the parameter space. At time $t=0$, the system starts in the ground state $\ket{G(\lambda_0)}$ with $\lambda_0 \equiv \lambda(t=0)$. During adiabatic evolution, where $\h(\lambda)$ varies slowly relative to the characteristic energy gaps, the system remains in the instantaneous ground state $\ket{G(\lambda)}$ and acquires a phase factor, as dictated by the adiabatic theorem~\cite{Born1928}. After an evolution time $T$, the Hamiltonian returns to its initial form, i.~e., $\h(\lambda_T) = \h(\lambda_0)$, with $\lambda_T \equiv \lambda(T)$. Without loss of generality, we assume that $\lambda\in [0, 2\pi]$ and that $\h$ is a periodic function of $\lambda$. By the end of the adiabatic cycle ($t = T$), the evolved quantum state can be expressed as:
\begin{align}
\ket{G(2\pi)} = \mathrm{e}^{-\mathrm{i}\varphi_\mathrm{D}} \mathrm{e}^{\mathrm{i}\varphi_\mathrm{B}} \ket{G(0)}\,,
\end{align}
where 
\begin{align}
 \varphi_\mathrm{D} = \int_0^T \mathrm{d}t\, E_\mathrm{G}[\lambda(t)]    
 \label{eq:phiD}
\end{align}
is the dynamical phase, which depends on the system's ground state energy $E_\mathrm{G}(\lambda)$ and the total evolution time $T$, and
\begin{align}
    \varphi_\mathrm{B} = -\mathrm{i} \int_0^{2\pi}\mathrm{d}\lambda\, \bra{G(\lambda)} \frac{\partial}{\partial \lambda} \ket{G(\lambda)}
\label{eq:phiB}
\end{align}
is the Berry phase, representing the geometric contribution to the total phase accumulated during the system's evolution in parameter space.

The foregoing discussion assumes adiabatic evolution along the closed parameter loop. Rather than invoking the traditional quantitative criterion, which is known to be neither sufficient nor necessary, we follow modern formulations that provide sufficiency and/or necessity together with rigorous error bounds. In finite-dimensional closed systems, gap-dependent sufficient conditions are available~\cite{Jansen2007,Albash2018}; the limitations of the older criterion and necessary variants have been analyzed~\cite{Tong2007,Comparat2009,Tong2010}; and gauge-invariant decompositions yield a necessary-and-sufficient framework with explicit nonadiabatic corrections~\cite{Wang2016}, with extensions to open or periodically driven settings~\cite{Santos2020,Gu2024}. In our dimerized Fermi-Hubbard chain models, $\h(\lambda)$ is smooth in $\lambda$ and the many-body gap remains open, so under the standard regularity hypotheses (smooth parameter dependence, a uniform minimum gap $\Delta E_{\mathrm min}>0$, and bounded time derivatives along the chosen schedule) the adiabatic error can be made arbitrarily small by increasing the total evolution time $T$. In practice, we corroborate this by monitoring the infidelity $1-f = 1 - |\langle G(\theta(\lambda))|G_{\mathrm{ED}}(\lambda)\rangle|^2$ along the loop and by verifying convergence of the extracted Berry phase at $\lambda=2\pi$ as $T$ grows; see Sec.~\ref{sec:res} for the numerical trends. We also note that exact asymptotic adiabaticity is not required for our protocol to return the correct Berry phase: the dynamical phase is canceled by time-reversed propagation, and the geometric phase is robust to moderate nonadiabatic admixtures in the regimes we study, as discussed below

\begin{figure*}[t!]

\begin{tikzpicture}
        \node[anchor=south west, inner sep=0] (image) at (0,0){};
        \node[above left] at (-11.5, 0) {(a)};
\end{tikzpicture}

\vspace{-5pt}
\centerline{\Qcircuit @C=1.0em @R=1.3em{ \lstick{|0\rangle } & \gate{H} &\ctrl{1} & \gate{H} &\meter\qw\\ \lstick{|\Psi_0\rangle} & \gate{U_G} & \gate{U_\text{loop}} &\qw & \qw & \qw 
}}

\begin{tikzpicture}
        \node[anchor=south west, inner sep=0] (image) at (0,0){};
        \node[above left] at (-11.5, 0) {(b)};
\end{tikzpicture}
    
\vspace{-5pt}
\centerline{\Qcircuit @C=1.0em @R=1.3em{ \lstick{|0\rangle } & \gate{H} &\ctrl{1}&\ctrl{1}&\ctrl{1} & \gate{H} &\meter\qw\\ \lstick{|\Psi_0\rangle} & \qw & \gate{U_\mathrm{G}[\bth^1(2\pi)]} & \gate{U_\text{loop}[\bth^2(2\pi)]} & \gate{U_\mathrm{G}[-\bth^1(0)]} & \qw & \qw & \qw 
}}

\caption{Hadamard test circuits for measuring the Berry phase.
(a) Circuit for Berry phase measurement using the cyclic adiabatic operator  $U_\text{loop}$. 
The ancillary qubit (upper horizontal line) is initialized in $\ket{0}$, with  Hadamard gates $H$ applied to it. The register for physical system qubits (lower horizontal line) is initially in the reference product state $\ket{\Psi_0}$. The unitary operator $U_\mathrm{G}$ prepares the ground state, $\ket{G(0)}=U_\mathrm{G}\ket{\Psi_0}$, while the ancilla-controlled application of  $U_\text{loop}$ enables cyclic adiabatic evolution. For the cyclic adiabatic evolution operator $U_\text{loop}$ defined in Eq.~\eqref{eq:Uloop}, measuring the ancilla qubit in the computational basis gives the probability $P_{\ket{0}}=\frac{1}{2}(1+\cos\varphi_\text{B})$ of observing the ancilla in state $\ket{0}$.  
(b) Berry phase measurement using a variational quantum circuit. The circuit employs parametrized unitaries $U_\mathrm{G}[\bth^1]$ and $U_\mathrm{loop}[\bth^2]$, generated via AVQITE and AVQDS, respectively. The angles of both parameterized unitaries evolve during the cyclic adiabatic evolution. After the adiabatic evolution, the circuit measures the overlap between the states $U_\mathrm{G}[\bth^1(0)]\ket{\Psi_0}$ and $U_\text{loop}[\bth^2(2\pi)]U_\mathrm{G}[\bth^1(2\pi)]\ket{\Psi_0}$. For sufficient accurate adiabatic evolution, these two states differ only by a phase $\varphi_\mathrm{qc}$, leading to the measurement probability $p_{\ket{0}} = \frac{1}{2} \left(1 + \cos \varphi_\mathrm{qc} \right)$.
Note that the practical implementation of the controlled-parameterized unitaries associated with the ancilla in (b) introduce only as many controlled single-qubit rotation gates as the number of variational parameters in the unitaries~\cite{Libbi2022}.
}
\label{fig1}    
\end{figure*}

\subsection{Quantum Circuit for Berry Phase Calculation}

To compute the Berry phase on a quantum computer, we follow the approach in Ref.~\citenum{Murta2020}, which employs a Hadamard test quantum circuit, illustrated in Fig.~\figref{fig1}{(a)}. This circuit measures the overlap between the initial and final states after an adiabatic loop.  Specifically, the ancilla qubit is first prepared in a superposition state, $\frac{1}{\sqrt{2}}(\ket{0} + \ket{1})$ via a Hadamard gate, while a unitary operator $U_\mathrm{G}$ initializes the system qubits in the ground state, $\ket{G(0)}=U_\mathrm{G}\ket{\Psi_0}$. A controlled-$U_{\text{loop}}$ operation is then applied, where $U_{\text{loop}}$ simulates the cyclic adiabatic evolution. The final Hadamard gate on the ancilla enables the measurement of the accumulated phase $\varphi$. Measuring the ancilla qubit in the computational basis yields the probability $P_{\ket{0}}=\frac{1}{2}(1+\cos\varphi)$ for observing the ancilla in state $\ket{0}$.

\begin{figure*}[t!]
\begin{center}
		\includegraphics[scale=0.5]{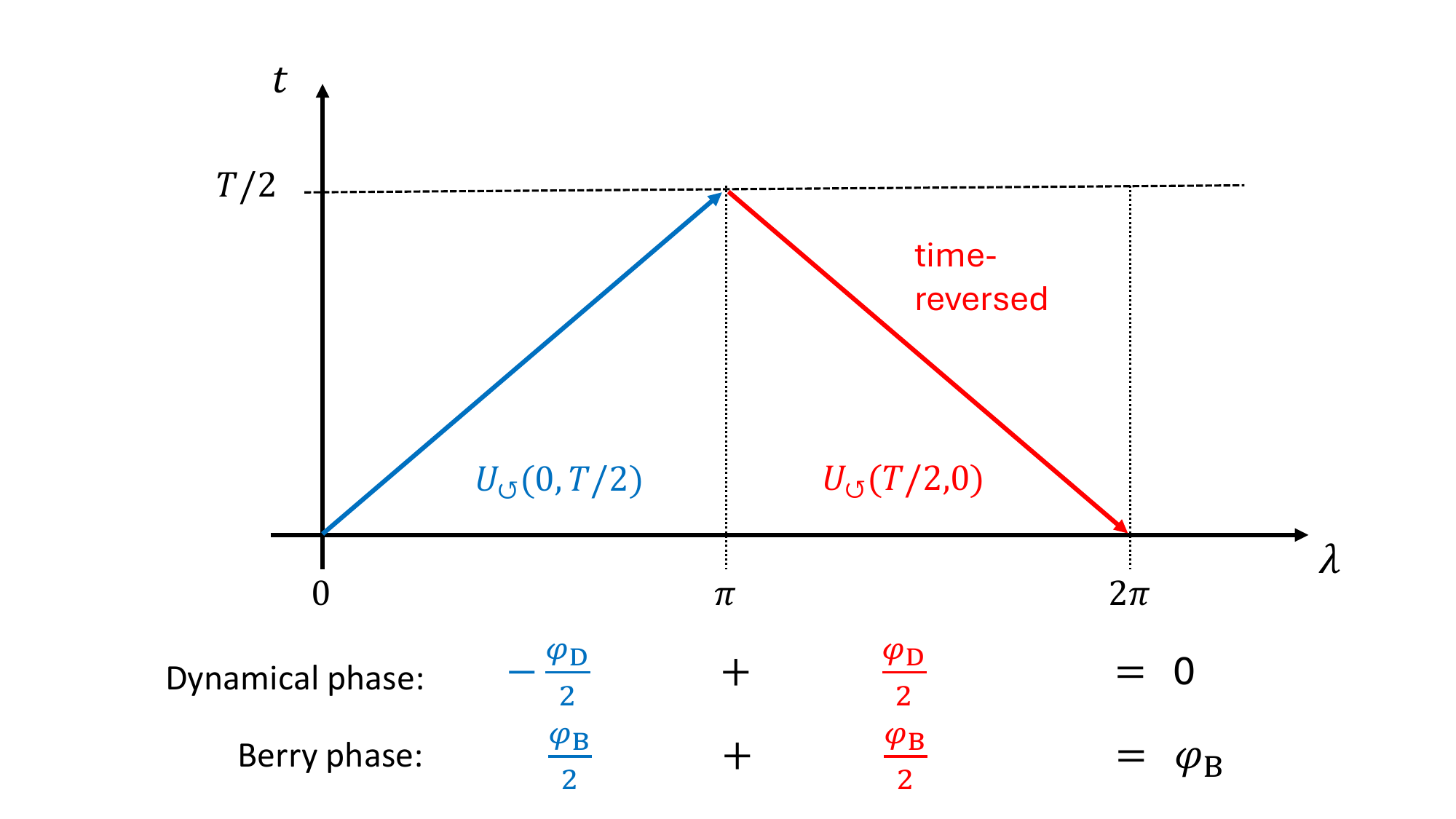}
\caption{Parameter-loop evolution illustrating dynamical-phase cancellation and Berry-phase addition. Forward half (blue arrow): $U_{\circlearrowleft}(0, T/2)$ with $t:0\to T/2$ and $\lambda:0\to\pi$. Return half (red arrow): $U_{\circlearrowleft}(T/2, 0)=U_{\circlearrowleft}(0,T/2)^\dagger$ executed with time-reversed propagation, $t:T/2\to 0$,  while $\lambda:\pi\to 2\pi$ in the same loop orientation. Annotations indicate $-\varphi_{\mathrm D}/2 + \varphi_{\mathrm D}/2 = 0$ (dynamical-phase cancellation) and $\varphi_{\mathrm B}/2 + \varphi_{\mathrm B}/2 = \varphi_{\mathrm B}$ (Berry-phase addition).}
		\label{fig0} 
\end{center}
\end{figure*}

To isolate the Berry phase $\varphi_\mathrm{B}$, the dynamical phase contribution is eliminated using time-reversal symmetry by choosing the evolution operator as
\begin{align}
U_{\text{loop}} = U_\circlearrowleft(0, T/2) U_\circlearrowleft(T/2, 0)\,.
\label{eq:Uloop}
\end{align}  
Here, the first (second) unitary represents forward (backward) evolution in time from $t=0$ to $t = T/2$ (from $t = T/2$ to $t=0$), while the parameter $\lambda$ evolves counterclockwise. Figure~\ref{fig0} schematically illustrates time $t$ versus the control parameter $\lambda$. The first half of the cycle ($U_{\circlearrowleft}(0, T/2)$, blue arrow) advances forward in time from 0 to $T/2$ while $\lambda$ increases from $0$ to $\pi$. The second half ($U_{\circlearrowleft}(T/2, 0)$, red arrow) uses the time-reversed propagator from $T/2$ back to 0 while $\lambda$ continues from $\pi$ to $2\pi$. For time-reversal-invariant Hamiltonians, the energy integral satisfies
\begin{align}
\int_0^\pi E_G(\lambda) \mathrm{d}\lambda = \int_\pi^{2\pi} E_G(\lambda) \mathrm{d}\lambda\,,
\end{align}  
ensuring the cancellation of the dynamical phase in the adiabatic loop defined by $U_\text{loop}$. Meanwhile, the Berry phase accumulates identically in both half-cycles, enabling direct computation of $\varphi_\mathrm{B}$ using the circuit in Fig.~\figref{fig1}{(a)}. The phase bookkeeping $-\varphi_{\mathrm D}/2+\varphi_{\mathrm D}/2=0$ and $\varphi_{\mathrm B}/2+\varphi_{\mathrm B}/2=\varphi_{\mathrm B}$ is illustrated in Fig.~\ref{fig0} for this symmetric case.

To implement the quantum circuit, the adiabatic evolution is discretized into $N$ steps, each lasting $\delta t$, during which $\lambda$ remains constant. The evolution operator can thus be expressed as  
\begin{align}
U_{\text{loop}} = \prod_{j=1}^{N} \delta U(\lambda_j, \delta t_j)\,,
\label{eq:Uloop-exp}
\end{align}  
where the unitary propagator is given by  
\begin{align}
\delta U(\lambda_j, \delta t_j) = \exp\left[-\mathrm{i} \h(\lambda_j) \delta t_j\right]\,,
\label{eq:deltaU}
\end{align}  
which can be implemented using Trotter decomposition~\cite{lloyd1996, Trotter_dynamics_Knolle,childs2018quamtumsimulation} as in Ref.~\citenum{Murta2020}. 
Here, $\delta t_j=\delta t (-\delta t$) in the first (second) half-cycle of the cyclic adiabatic evolution  with a total duration $T=N \delta t$. The discretized time is defined as $t_j = \sum_{k=1}^{j} \delta t_k$ with $t_0 = t_N = 0$. In the first half-cycle, $t_j$ evolves from $0$ to $T/2$ while $\lambda$ increases from $0$ to $\pi$ such that $\lambda_j = \frac{2\pi}{T} t_j$. In the second half-cycle, $t_j$ evolves from $T/2$ to $0$ and $\lambda$ increases from $\pi$ to $2\pi$, yielding  $\lambda_j = \frac{2\pi}{T} (T - t_j)$. 

To achieve an adiabatic state evolution, the total cyclic evolution time $T$, which determines the quench rate, must be sufficiently large to ensure that the Hamiltonian evolves slowly relative to the minimal energy gap along the path, allowing the system to remain in the instantaneous ground state. Meanwhile, the time step size $\delta t$ should be small enough to maintain a high accuracy of the numerical dynamics simulation.

\subsection{Variational quantum circuit for Berry phase measurement}

Instead of using Trotter decomposition and the quantum circuit in Fig.~\figref{fig1}{(a)}, we adopt the AVQDS approach to automatically construct compact quantum circuits for Berry phase estimation. The corresponding AVQDS-based circuit is shown in Fig.~\figref{fig1}{(b)}. First, we use the adaptive variational quantum imaginary time evolution (AVQITE) algorithm~\cite{AVQITE} to prepare the system's ground state, $\ket{G[\bth^1(\lambda=0)]} = U_\mathrm{G}[\bth^1(0)]\ket{\Psi_0}$, where $U_\mathrm{G}[\bth^1(0)]$ represents a sequence of parameterized unitaries applied to a reference product state $\ket{\Psi_0}$ with $\lambda=0$. The adiabatic evolution is then implemented using additional unitaries $U_\text{loop}[\bth^2]$, which are dynamically generated by the AVQDS algorithm.  Importantly, both $\bth^1$ and $\bth^2$ evolve in time according to Eq.~\eqref{eq:theta_eom} for leveraging the variational degrees of freedom of the entire ansatz, yielding a highly compressed circuit while maintaining accuracy~\cite{mootz2024adaptive}. 

The quantum circuit in Fig.~\figref{fig1}{(b)} measures the overlap between the initial state, $U_\mathrm{G}[\bth^1(0)]\ket{\Psi_0}$, and final state of the adiabatic loop, $U_\text{loop}[\bth^2(2\pi)]U_\mathrm{G}[\bth^1(2\pi)]\ket{\Psi_0}$, which can be derived from the probability  $p_{\ket{0}}=\frac{1}{2} + \frac{1}{2}\mathrm{Re}\left[\bra{\Psi_0} U_\mathrm{G}[-\bth^1(0)] U_\text{loop}[\bth^2(2\pi)]U_\mathrm{G}[\bth^1(2\pi)]\ket{\Psi_0}\right]$. For sufficiently accurate adiabatic evolution, the states differ only by a phase $\mathrm{e}^{\mathrm{i} \varphi_\mathrm{qc}}$,
such that $p_{\ket{0}} = \frac{1}{2} \left(1 + \cos \varphi_\mathrm{qc} \right)$. As discussed in Sec.~\ref{sec:avqds}, our implementation of AVQDS does not explicitly track the global phase. As a result, we account for it by adding $\varphi_\mathrm{G}$, determined from Eq.~\eqref{eq:phiG}, to $\varphi_\mathrm{qc}$ to obtain the Berry phase: 
\begin{align}  
\varphi_\mathrm{B} = \varphi_\mathrm{qc} + \varphi_\mathrm{G}\,.  
\label{eq:phiB2}
\end{align}  

The right-hand side of Eq.~\eqref{eq:phiG} consists of contributions related to the dynamical phase and geometric properties of the cyclic adiabatic evolution. Integrating both terms separately over time gives
\begin{align}
    &\varphi_\mathrm{G}(t) = \varphi_\mathrm{G,1}(t) + \varphi_\mathrm{G,2}(t)\,, \nonumber \\
    &\varphi_\mathrm{G,1}(t)= -\int_0^t\mathrm{d}t'\, \langle \h \rangle_{\bth} \,,\nonumber \\
    &\varphi_\mathrm{G,2}(t)= \int_0^t\mathrm{d}t'\,  \mathrm{Im} \left[\sum_\mu \langle\Psi[\bth]|\frac{\partial|\Psi[\bth]\rangle}{\partial \theta_\mu} \frac{\mathrm{d}\theta_\mu}{\mathrm{d}t'} \right]\,.
    \label{eq:phiGi}
\end{align}
Here, $\varphi_\mathrm{G,1}$, corresponds to the dynamical phase, as defined in Eq.~\eqref{eq:phiD}, while $\varphi_\mathrm{G,2}$ captures the geometric properties of the adiabatic evolution. As shown in Appendix~\ref{sec:app}, for numerically exact cyclic adiabatic quantum simulations in the adiabatic limit $T \to \infty$ and for time-reversal symmetric variational parameters, both global phase contributions $\varphi_\mathrm{G,1}$ and $\varphi_\mathrm{G,2}$ vanish. Consequently, the quantum circuit in Fig.~\figref{fig1}{(b)} then directly yields the Berry phase. For finite $T$ and due to the adaptive nature of AVQDS, $\varphi_\mathrm{G}$ is generally nonzero. Nevertheless, in practice, we find a perfect error cancellation between $\varphi_\mathrm{G}$ and $\varphi_\mathrm{qc}$ such that $\varphi_\mathrm{B} = \varphi_\mathrm{G}+\varphi_\mathrm{qc}$ remains robust even for small $T$ with strong nonadiabaticity.

\section{Results \label{sec:res}}

\subsection{Model}

To benchmark our approach for calculating the Berry phase, we apply it to the dimerized Fermi-Hubbard model, also known as the Su–Schrieffer–Heeger–Hubbard (SSHH) model~\cite{Ye2016,Le2020,Mikhail2024,Chang:2024ttv}. This model provides a versatile platform for exploring strongly correlated quantum systems and their topological properties, particularly the interplay between electron interactions and lattice dimerization.
We consider the one-dimensional SSHH model with $N$ sites and periodic boundary conditions, governed by the Hamiltonian:
\begin{align}
    \h &= \sum_{j, \sigma} \left[ t_{j,j+1} \hat{c}_{j+1,\sigma}^\dagger \hat{c}_{j,\sigma} + \text{h.c.} \right] \nonumber \\
    &+ U \sum_j \hat{n}_{j, \uparrow} \hat{n}_{j, \downarrow} - \frac{U}{2} \sum_{j, \sigma} \hat{n}_{j, \sigma}\,.
    \label{eq:Ham}
\end{align}
Here, $t_{j,j+1} = t(1 + (-1)^j \delta)$ represent the staggered hopping amplitudes with dimerization parameter $\delta$, $U$ is the on-site interaction strength, $\hat{c}_{j,\sigma}^\dagger$ and $\hat{c}_{j,\sigma}$ are the creation and annihilation operators for a fermion with spin $\sigma$ at site $j$, and $\hat{n}_{j,\sigma} = \hat{c}_{j,\sigma}^\dagger \hat{c}_{j,\sigma}$ is the number operator. The third term $-\frac{U}{2} \sum_{j, \sigma} \hat{n}_{j, \sigma}$ ensures particle-hole symmetry. For $U=0$, the model reduces to the noninteracting Su-Schrieffer-Heeger (SSH) model, a prototypical topological insulator characterized by a quantized Berry phase that changes when $\delta$ switches sign~\cite{Murta2020}. In the strong interaction limit $U/t \to \infty$, the model maps onto a spin-$1/2$ Heisenberg chain with alternating exchange couplings. In our numerical simulations, we focus on a $N=4$-site chain, set the hopping parameter to $t=1$, and study cases with $U=0$ and $U=10$ in the noninteracting and strongly correlated regimes, respectively. 

To compute the Berry phase, we impose twisted boundary conditions~\cite{Xiao2010, TBC} selectively on spin-up fermions:
\begin{align}
    \hat{c}_{j+N, \uparrow} = e^{i\lambda} \hat{c}_{j, \uparrow}\,,
\end{align}
where $\lambda$ is the twist angle. Spin-down fermions remain unaffected.  This selective twisted boundary condition application is crucial since applying it to both spin species constrains the Berry phase to integer multiples of $2\pi$, making it unsuitable for tracking topological phase transitions. To calculate the Berry phase using the algorithm in Sec.~\ref{sec:methods}, we perform a cyclic adiabatic evolution by continuously varying the twist angle $\lambda$ from 0 to $2\pi$.

To simulate the SSHH model on a quantum computer, the Hamiltonian must be mapped to a multi-qubit operator. Following the approach in our previous work~\cite{mootz2024adaptive}, we utilize the Jordan-Wigner transformation~\cite{map_jw} to encode the fermionic creation and annihilation operators into qubit operators. For a $N=4$ site system, this encoding requires $N_q = 8$ qubits.

\subsection{Ground state preparation}

We prepare the ground state using the Adaptive Variational Quantum Imaginary Time Evolution (AVQITE) algorithm~\cite{AVQITE}, which iteratively refines the variational ansatz by minimizing its deviation from the exact imaginary time-evolved state, following McLachlan’s variational principle~\cite{mclachlan64variational}. Similar to AVQDS, this method adaptively expands the ansatz to keep the McLachlan distance below a predefined threshold, selecting unitaries from a predetermined operator pool. The resulting ansatz retains a pseudo-Trotter structure, as expressed in Eq.~\eqref{eq:ansatz}, ensuring compatibility with the AVQDS approach.

Following Ref.~\citenum{mootz2024adaptive}, we construct the operator pool derived from unitary coupled-cluster singles and doubles excitations~\cite{MayhallQubitAVQE, smqite} or their equivalent qubit excitation operators~\cite{yordanov2021qubit}. The pool consists of:
\begin{align}
    \mathcal{P} = \{ \hat{\sigma}_i^p\hat{\sigma}_j^q\} \cup \{ \hat{\sigma}_i^p\hat{\sigma}_j^q \hat{\sigma}_k^r \hat{\sigma}_l^s\}\,,
    \label{eq:sd_pool}
\end{align}
where only Pauli strings containing an odd number of $\hat{\sigma}^y$ operators are included. The indices $i,j,k,l$ label qubits and $p, q, r, s \in \{x, y\}$. For a system of size $N=4$, this approach yields a pool of $N_\mathrm{p} = 616$ generators.

As the reference state  $\ket{\Psi_0}$, we use a simple product state in which spin-up electrons occupy the first $N/2$ fermionic orbitals while spin-down electrons occupy the remaining $N/2$. This state is then mapped into its qubit representation.

\begin{figure*}[t!]
\begin{center}
		\includegraphics[width=0.87\textwidth]{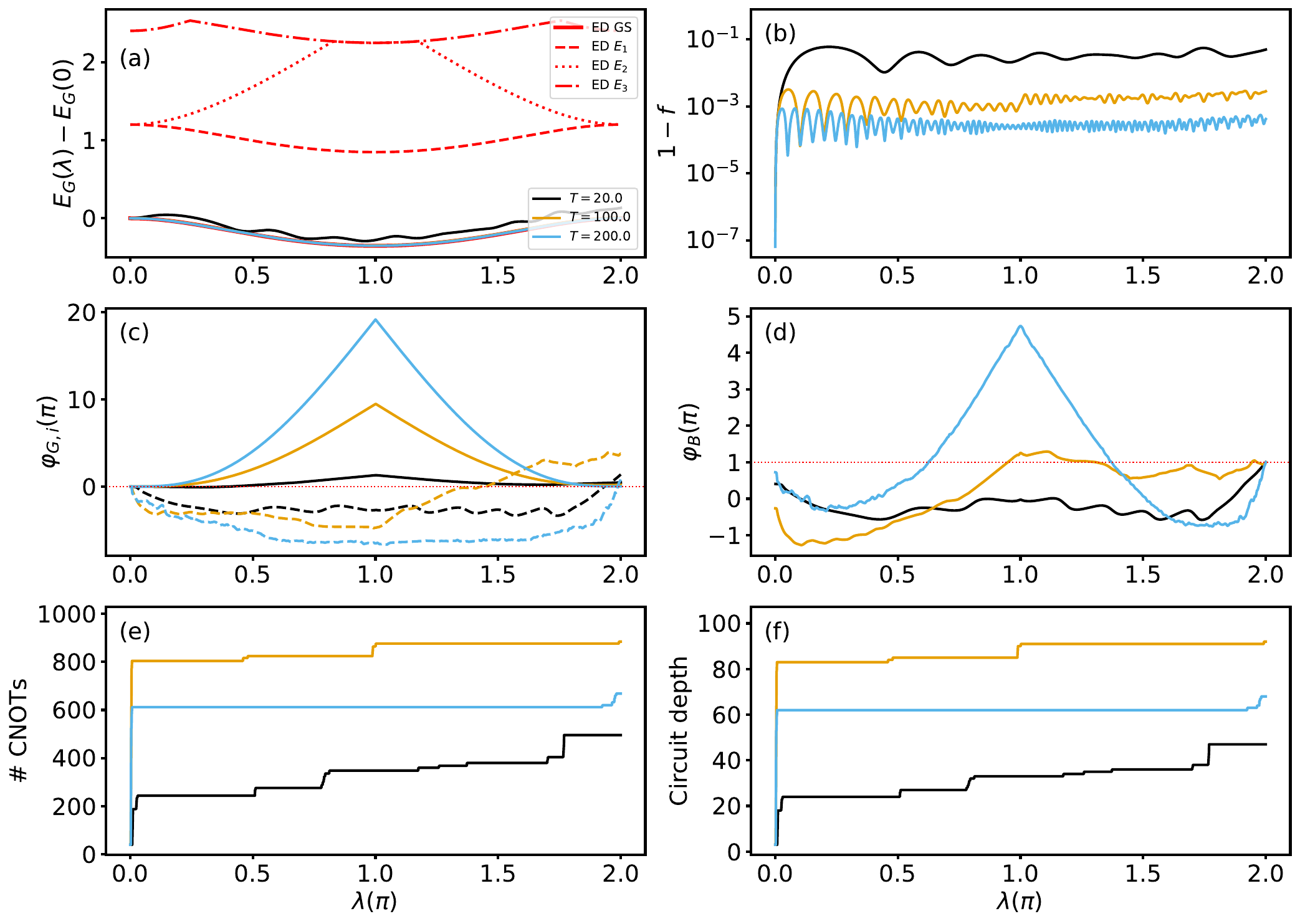}
		\caption{Benchmarking of the AVQDS approach for calculating the Berry phase of the noninteracting SSH model. (a) The energy $E(\lambda)$ of the quantum state evolved by AVQDS for $T = 20$ (black line), $T = 100$ (orange line), and $T = 200$ (blue line), compared to exact diagonalization (ED) results (red line), for a dimerization parameter of $\delta=-0.3$.  For comparison, the first three ED excited energies $E_{1,2,3}(\lambda)$ are plotted in red with distinct dashed line styles. (b) Corresponding infidelity $1 - f$, showing improved accuracy with increasing $T$. The maximum infidelity across all $\lambda$ is $5.0\times 10^{-2}$ for $T=20$, $2.8\times 10^{-3}$ for $T=100$, and $4.2\times 10^{-4}$ for $T=200$. (c) Contributions to the global phase $\varphi_\mathrm{G}$: dynamical phase $\varphi_{\mathrm{G},1}(\lambda)$ (solid lines) and geometric phase contribution $\varphi_{\mathrm{G},2}(\lambda)$ (dashed lines), calculated by solving Eq.~\eqref{eq:phiG} using a fourth-order Runge-Kutta method. The red dashed line marks $\varphi_{\mathrm{G},i} = 0$. (d) Comparison of the Berry phase computed via Eq.~\eqref{eq:phiB2} with $\varphi_\mathrm{B}$ obtained from ED (red dashed line), demonstrating the robustness of the AVQDS approach. The Berry phase should be read at the end of the loop, $\lambda=2\pi$, where the curves for different $T$ converge, rather than across the entire $\lambda$ range. (e) Number of CNOT gates as a function of $\lambda$ with maximum counts of 496, 884, and 668 for $T = 20$, $T=100$, and $T = 200$, respectively. (f) Circuit depth (number of layers of unitaries) during cyclic state evolution, reaching 47 layers for $T = 20$, 92 layers for $T = 100$, and 68 layers for $T = 200$.} 
		\label{fig2} 
\end{center}
\end{figure*}

\subsection{Simulation results}

\subsubsection{Noninteracting SSH Model}

We first evaluate the performance and quantum circuit complexity of the algorithm outlined in Sec.~\ref{sec:methods} for the noninteracting SSH model ($U=0$). In these calculations, we employ the Hamiltonian operator pool for AVQDS, which consists of all individual Pauli strings in the qubit representation of the Hamiltonian. For $N=4$ sites this pool contains $N_\mathrm{p}=18$ distinct operators.
Figure~\figref{fig2}{(a)} presents the ground-state energy $E_\mathrm{G}(\lambda)$ obtained using the AVQDS approach with evolution times $T = 20$ (black line), $T = 100$ (orange line), and $T = 200$ (blue line), compared to the adiabatic state result from exact diagonalization of the time-dependent Hamiltonian with $T=200$ (red line). The results are presented for a dimerization parameter $\delta = -0.3$. As expected, increasing $T$ improves the accuracy of $E_\mathrm{G}(\lambda)$, with significant deviations from the exact energy observed for $T = 20$ due to substantial nonadiabatic effects.
To quantify the nonadiabaticity of the state evolution due to a finite $T$, we evaluate the infidelity $1-f=1-|\ov{G(\bth(\lambda))}{ G_\mathrm{ED}(\lambda)}|^2$, where $\ket{G(\bth(\lambda))}$ is the AVQDS-propagated state and $\ket{G_\mathrm{ED}(\lambda)}$ is the instantaneous ground state obtained by exact diagonalization of the Hamiltonian $\h(\lambda)$.
Figure~\figref{fig2}{(b)} shows $1-f$ for the same values of $T$ used in Fig.~\figref{fig2}{(a)}. For $T = 20$, the peak value of the infidelity is $0.05$, indicating a noticeable deviation from the exact ground state due to nonadiabaticity. This imperfect state evolution leads to the large energy inaccuracy observed in Fig.~\figref{fig2}{(a)}. As $T$ increases, the infidelity decreases, remaining below $5\times 10^{-4}$ for $T = 200$ throughout the entire cyclic adiabatic evolution. 

The oscillations of $1-f$ in Fig.~\figref{fig2}{(b)} originate from a small diabatic admixture of low-lying excited states induced by the finite ramp speed: ground- and excited-state components acquire different dynamical phases whose interference produces oscillations as $\lambda$ advances. For a linear ramp $\lambda(t)=2\pi t/T$ and a constant lowest gap $\Delta E_1 = E_1 - E_0 = 1.2$ (Fig.~\figref{fig2}{(a)}), the $\lambda$-domain angular frequency is $\Omega_\lambda= \Delta E_1 T/(2\pi)$, implying a period that scales as $1/T$. Numerically, $\Omega_\lambda\approx 3.8,\,19.1,\,38.2$ such that one oscillation spans $\Delta\lambda/\pi=2/\Omega_\lambda\approx 0.52,\,0.10,\,0.05$ for $T=20,100,200$, consistent with the small-$\lambda$ oscillations in Fig.~\figref{fig2}{(b)}. At larger $\lambda$, higher excited states admix and the instantaneous frequency varies accordingly.

To corroborate that the infidelity $1 - f$ characterizes the nonadiabaticity of the cyclic state evolution with a finite $T$, we examine the accuracy of AVQDS simulations by monitoring the infidelity, $1-f_t=1-|\ov{G(\bth(\lambda))}{ G_\mathrm{ED, t}(\lambda)}|^2$, where
$\ket{G_\mathrm{ED, t}(\lambda)}$ is evolved using ED (Eq.~\eqref{eq:ED}) with the same $T$. This metric quantifies the accuracy of AVQDS in capturing the quantum dynamics. We find that the maximum value of $1-f_t$ during the full evolution is $1.1\times 10^{-4}$ for $T=20$ while it is $1.3\times 10^{-4}$  ($1.4\times 10^{-4}$ ) for $T=100$ ($200$). These values are consistently smaller than the corresponding $1-f$, confirming that $1-f$ is dominated by nonadiabatic effects due to a finite $T$.

Next, we analyze the contributions to the Berry phase $\varphi_\mathrm{B}$ arising from the global phase determined by the equation of motion~\eqref{eq:phiG}. Figure~\figref{fig2}{(c)} illustrates the components of the global phase $\varphi_\mathrm{G}$, as defined in Eq.~\eqref{eq:phiGi}. The dynamical phase component, $\varphi_{\mathrm{G},1}(\lambda)$, is shown as solid lines, while the geometric phase component, $\varphi_{\mathrm{G},2}(\lambda)$, is depicted as dashed lines. The dynamics of these components are computed by solving the equation of motion~\eqref{eq:phiG} using a fourth-order Runge-Kutta method, treating the two terms on the right-hand side separately.  
Notably, for all studied values of $T$, the dynamical phase $\varphi_{\mathrm{G},1}$ is close to zero, indicating accurate cancellation during cyclic adiabatic evolution. In contrast, the geometric component $\varphi_{\mathrm G,2}$ decreases only gradually with $T$ and does not follow a simple ordering $T=20 > T=100 > T=200$. As shown in Appendix~\ref{sec:app}, for a time-reversal-invariant Hamiltonian the forward and return contributions, $\varphi_{\mathrm G,2}^{+}$ and $\varphi_{\mathrm G,2}^{-}$, cancel if the return half uses the same ansatz and follows the time-reversed parameter path, i.~e., $\bth^{-}(\lambda)=\bth^{+}(2\pi-\lambda)$ $(\lambda:\pi\to 2\pi)$. In AVQDS this symmetry is only approximate and depends not on $T$ alone but on where the adaptive ansatz grows along the loop. For sufficiently large $T$, most operator insertions occur in the first half, the return half effectively reuses a fixed ansatz, and the cancellation is nearly exact, driving $\varphi_{\mathrm G,2}\!\to\!0$. For intermediate $T$, however, insertions in the middle or in the return half break the symmetry and produce larger residuals. This mechanism accounts for the non-monotonic behavior and the ordering observed in Fig.~\figref{fig2}{(c)}.

Figure~\figref{fig2}{(d)} presents the resulting Berry phase $\varphi_\mathrm{B}$, obtained using Eq.~\eqref{eq:phiB2}. Note the Berry phase should be read at the loop endpoint $\lambda=2\pi$, where the dependence on $T$ vanishes and the curves converge. Here, the phase $\varphi_\mathrm{B}=\varphi_\mathrm{qc}+\varphi_\mathrm{G}$ is the sum of $\varphi_\mathrm{qc}$ (a constant) computed with the circuit in Fig.~\figref{fig1}{(b)} and the global phase $\varphi_\mathrm{G}$ determined via Eq.~\eqref{eq:phiG}. The red dashed line represents the Berry phase extracted from ED. Despite the nonadiabatic effects observed for $T = 20$ and $T = 100$ in Figs.~\figref{fig2}{(a-c)}, the computed Berry phase remains highly accurate, underscoring the robustness of the AVQDS approach to calculate topological properties of quantum states. A more detailed analysis of this robustness is provided below.

Although the mid-path fidelities in Fig.~\figref{fig2}{(b)} are only moderate, the Berry phase in Fig.~\figref{fig2}{(d)} remains accurate because it is a geometric, gauge-invariant quantity (defined modulo $2\pi$) determined by the entire closed, gapped loop rather than by the state at any single point along the path. Expanding $|\psi(\lambda)\rangle$ in the instantaneous eigenbasis $\{|G(\lambda)\rangle,|m(\lambda)\rangle_{m>0}\}$ of $\h(\lambda)$ with eigenvalues $E_0(\lambda)$ and $E_m(\lambda)$ (energy gaps $\Delta E_m(\lambda)=E_m(\lambda)-E_0(\lambda)$), we obtain $|\psi(\lambda)\rangle=\alpha(\lambda)\,|G(\lambda)\rangle+\sum_{m>0}\beta_m(\lambda)\,|m(\lambda)\rangle$. The fidelity with the instantaneous ground state is then $F(\lambda)=|\alpha(\lambda)|^2$ while the infidelity is $1-F(\lambda)=\sum_{m>0}|\beta_m(\lambda)|^2$. The excited-state components $\beta_m(\lambda)$ carry rapidly varying dynamical phases set by the gaps. Our forward segment followed by its time-reversed return cancels the dynamical contribution exactly, so the remaining phase is geometric and is governed by the smoothly transported ground-state component as long as $|\alpha(\lambda)|>0$ along the path and the gap does not close. Over the closed loop, the residual contributions from the small excited admixtures enter with fast, oscillatory phases and largely average out, which explains why the Berry phase at the end of the loop can converge even when mid-path fidelities are only moderate. Consistent with this mechanism and prior robustness studies of geometric phases~\cite{Zhang2017,Colmenar2022}, we observe convergence of $\varphi_\mathrm{B}$ with increasing $T$ and mesh refinement.

\begin{figure*}[t!]
   \centering
    \includegraphics[width=\textwidth]{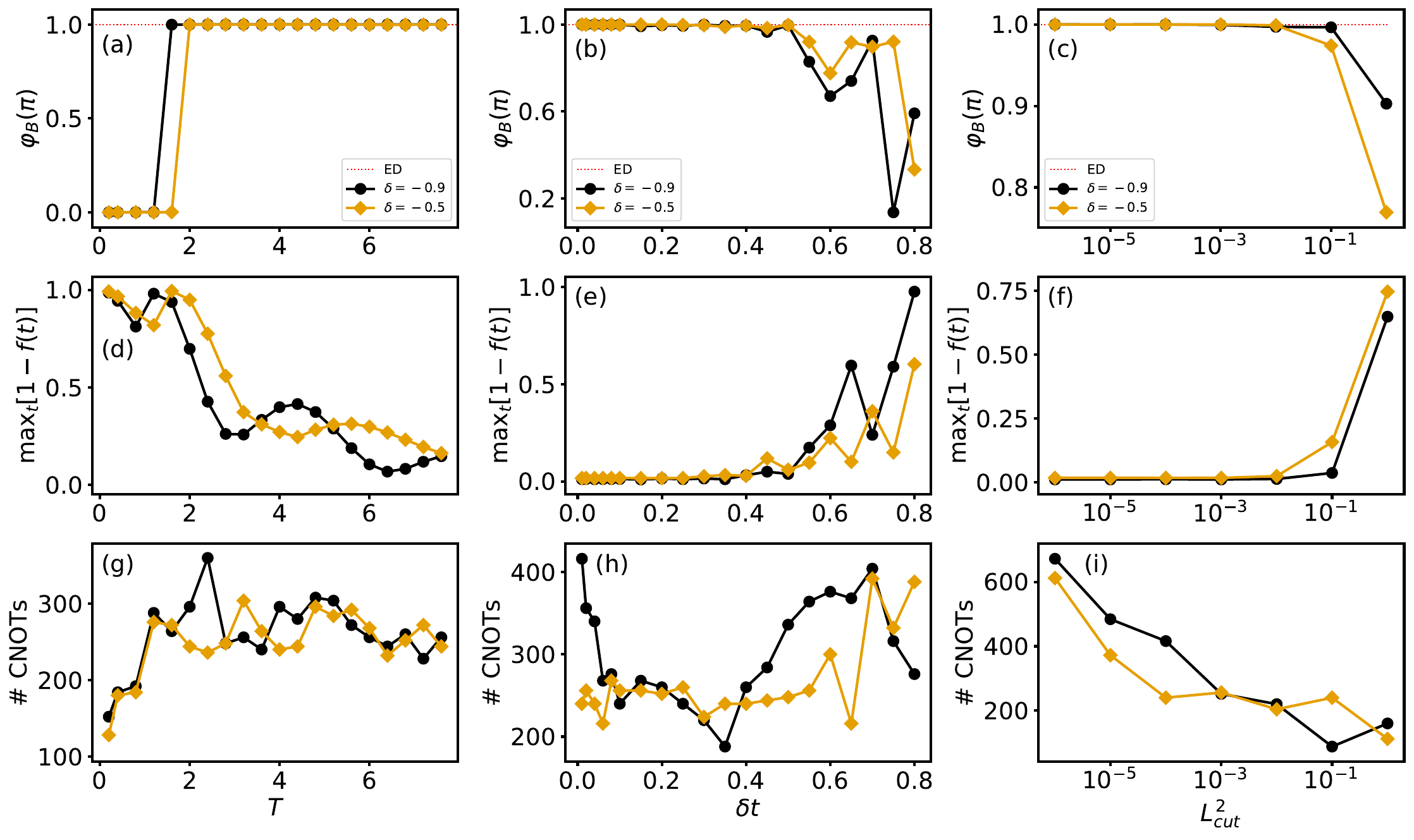}
    \caption{Robustness of the AVQDS algorithm in Berry phase calculations.
    (a) Computed Berry phase $\varphi_\mathrm{B}$ as a function of the cyclic state evolution time $T$ for $\delta = -0.9$ (black circles) and $\delta = -0.5$ (orange diamonds). The ED result for $T=200$ is shown as a red dashed line for reference. Accurate results can already be obtained for small $T = 1.6$ ($T = 2.0$) for $\delta=-0.9$ ($\delta=-0.5$), despite strong nonadiabaticity in this regime.
    (b) Computed Berry phase as a function of time step size $\delta t$ using a fixed $T = 20$ for $\delta = -0.9$ (black circles) and $\delta = -0.5$ (orange diamonds), compared with the ED result (red dashed line). The Berry phase is accurately obtained for $\delta t \leq 0.5$, but significant deviations appear beyond this threshold. 
    (c) Computed Berry phase as a function of McLachlan distance cutoff $L^2_\text{cut}$ using a fixed $T = 20$ for $\delta = -0.9$ (black circles) and $\delta = -0.5$ (orange diamonds), alongside the ED result (red dashed line). Accurate results are maintained for $L^2_\text{cut} \leq 10^{-1}$ ($10^{-2}$) for $\delta=-0.9$ ($\delta=-0.5$) before strong deviations emerge.
    (d-f) Corresponding maximum infidelities during the cyclic state evolutions, $\max_t[1 - f(t)]$, for the data in panels (a-c). Apart from the $T$-dependence, the infidelity remains low below the critical parameter values but increases significantly above them.
    (g-i) Corresponding CNOT gate counts for the data in panels (a-c). The CNOT count remains stable with increasing $T$ (g), decreases with increasing $\delta t$ for $\delta t=-0.9$ but remains stable for $\delta t=-0.5$ until the accuracy threshold (h), and decreases by a factor of 4 (3) for $\delta=-0.9$ ($\delta=-0.5$) as $L^2_\text{cut}$ increases to its critical value (i).
    }
    \label{fig3}
\end{figure*}

All results presented here are obtained using noiseless state-vector simulations, neglecting errors from quantum gates and finite sampling when measuring the matrix $M$ and vector $V$ in Eqs.~\eqref{eq:M} and \eqref{eq:V}. Nevertheless, the simulations allow straightforward analysis of the quantum circuit complexity. Specifically, we analyze the number of required two-qubit CNOT gates and the circuit depth, defined as the number of layers of parameterized unitaries. For estimating the CNOT count, we assume all-to-all qubit connectivity, as available in trapped-ion architectures, where implementing a unitary $e^{-i\theta \hat{A}}$ with a weight-$p$ Pauli string $\hat{A}$ requires $2(p-1)$ CNOT gates.
Figure~\figref{fig2}{(e)} shows the number of CNOT gates as a function of $\lambda$ for the three different values of $T$. The AVQITE ground-state preparation requires 20 CNOT gates, achieving an infidelity of $2.0\times 10^{-8}$. The number of CNOT gates increases primarily during the first half cycle of the adiabatic evolution, reaching 496 at $\lambda = 2\pi$ for $T = 20$. The peak CNOT count increases to 814 for $T = 100$ and 668 for $T = 200$. Interestingly, the CNOT count for $T = 100$ exceeds that for $T = 200$, indicating a nontrivial scaling of quantum resource requirements with evolution time.

Figure~\figref{fig2}{(f)} presents the corresponding circuit depth, measured in terms of the number of layers of unitaries. For $T = 20$, the circuit depth reaches 47 layers, while it increases to 92 layers for $T = 100$ and 68 layers for $T = 200$. Note that the circuit depths are comparable to those implemented on current quantum hardware. For example, quantum utility has been showcased with 20 Trotter steps or 60 CNOT layers for the dynamics of transverse-field Ising model in Ref.~\citenum{kimEvidenceUtilityQuantum2023}, and with 90 Trotter steps or 180 CNOT layers for energy transport studies of mixed-field Ising model in Ref.~\citenum{chen2023problemts}.

To contextualize the resource savings of our approach, we compare against first-order Trotterization. The time-dependent Hamiltonian decomposes into 18 Pauli strings (12 of weight 3 and 6 of weight 7). Assuming full connectivity and a compilation cost of $2(k-1)$ CNOTs for a weight-$k$ string, the CNOTs per Trotter step are $N_{\mathrm{CNOT/step}}=120$. With a uniform step size $\delta t=0.005$ (chosen to match the AVQDS accuracy), the total CNOT count is $N_{\mathrm{CNOT}}^{\mathrm{Trotter}}(T)=120\,T/0.005$,
i.~e., $4.80\times 10^{5}$, $2.40\times 10^{6}$, and $4.80\times 10^{6}$ for $T=20,100,200$, respectively. This implies that first-order Trotterization requires $\sim 9.7\times 10^{2}$, $\sim 2.7\times 10^{3}$, and $\sim 7.2\times 10^{3}$ more CNOTs than AVQDS for $T=20,100,200$, respectively (see Fig.~\figref{fig2}{(e)} for the AVQDS counts). In terms of circuit depth, using the AVQDS peak depths of 47, 92, and 68 layers for $T=20,100,200$ (Fig.~\figref{fig2}{(f)}), the corresponding Trotter/AVQDS depth ratios under the same serialized compilation and connectivity assumptions are approximately $\{1.02\times 10^{4},\,2.61\times 10^{4},\,7.06\times 10^{4}\}$. These estimates underscore the substantial resource compression achieved by AVQDS at matched accuracy and are consistent with earlier results~\cite{AVQDS,mootz2023twodimensional}.

It is important to note that the CNOT and depth trends in Fig.~\figref{fig2}{(e,f)} do not exhibit a simple dependence on $T$ but instead reflect the adaptive nature of AVQDS. Changing $T$ alters the quench rate along the $\lambda$-loop and, consequently, the pattern of ansatz growth. Three coupled factors govern the resource totals: (i) how many operator insertions are triggered (very small $T$ leaves little opportunity; very large $T$ is often captured by a few early insertions; intermediate $T$ can reveal additional segments that benefit from updates), (ii) where along the loop the insertions occur (additions near the start or end allow the return half to reuse the same ansatz, whereas mid-loop insertions increase totals), and (iii) the order in which operators are appended (early choices can reshape later needs). Together these effects yield higher counts at $T=100$ than at $T=200$ in the noninteracting case, while in the interacting case, discussed below, the same mechanisms act at different $\lambda$ locations, producing $T=200 > T=20 > T=100$ in Fig.~\figref{fig5}{(f)}.

While Figs.~\figref{fig2}{(e)}, \figref{fig2}{(f)} report concrete CNOT/depth profiles for a fixed system size across several evolution times, it is helpful to place these numbers in a broader algorithmic context. The quantum resource requirement for the Berry-phase calculations includes (i) ground-state preparation and (ii) cyclic adiabatic evolution. Aiming for near-term applications, we employ adaptive variational methods that automatically generate compact, problem-specific ans\"atze: AVQITE for ground-state preparation and AVQDS for adiabatic time evolution along the closed parameter loop. Because the circuits are selected adaptively and benchmarks often cover only a limited range of system sizes, the determination of the exact scaling with the number of sites is challenging. Moreover, the circuit depth is model- and problem-dependent due to the heuristic and variational nature. Nevertheless, several numerical benchmarks exist. For AVQITE, highly compact ground-state ans\"atze with remarkably lower depth than problem-agnostic unitary coupled-cluster (UCC) ansatz and Hamiltonian variational ansatz have been demonstrated in molecular and lattice settings~\cite{AVQITE}. For AVQDS, the system-size scaling of two-qubit resources (e.~g., CNOT count) has been investigated across representative spin and lattice models (including higher-spin cases)~\cite{AVQDS,mootz2023twodimensional}; while finite sizes preclude a definitive asymptotic bound, these benchmarks consistently indicate polynomial growth of circuit depth with system size on the tested models and sizes. Adaptive variational algorithms naturally leverage QPU strengths through compact representations of many-body states and unitary time evolution, making them promising for scaling Berry-phase simulations toward regimes beyond classical reach.

\begin{figure*}[t!]
    \centering
    \includegraphics[width=\textwidth]{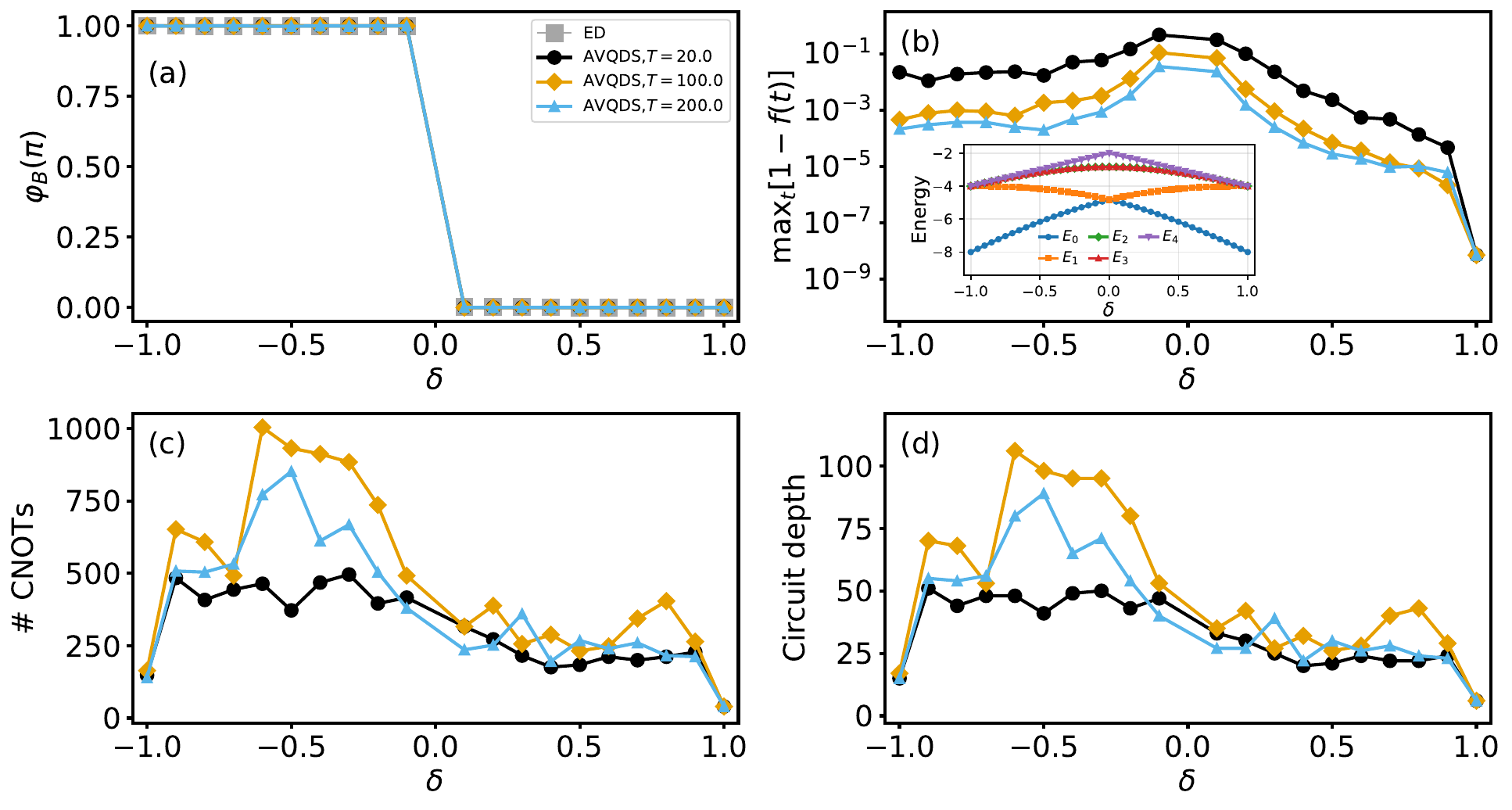}
    \caption{Quantum circuit complexity and fidelity for Berry phase calculations across a topological phase transition in the noninteracting SSH model.
    (a) Computed Berry phase $\varphi_\mathrm{B}$ as a function of the dimerization parameter $\delta$ for evolution times $T = 20$ (black circles), $T = 100$ (orange diamonds), and $T = 200$ (blue triangles), compared to exact diagonalization (ED) results (gray squares). All AVQDS calculations accurately reproduce the ED results and resolve the phase transition at $\delta = 0$.
    (b) Maximum infidelity during the cyclic adiabatic evolution as a function of $\delta$ for the three values of $T$. The infidelity peaks near the phase transition at $\delta = 0$ from both sides and decreases with increasing $T$, staying below $3.6\times 10^{-2}$ for $T = 200$. Inset: Lowest five many-body energies $E_{0\ldots4}$ vs. dimerization $\delta$ under a twisted boundary with $\lambda=\pi$. The gap $\Delta=E_1-E_0$ narrows as $\delta\!\to\!0$, explaining the rise of nonadiabatic effects and the need for longer evolution times near the transition at $\delta=0$.
    (c) Number of CNOT gates as a function of $\delta$ for the three different $T$. The highest CNOT count occurs for $T = 100$ in the nontrivial topological region ($\varphi_\mathrm{B} = \pi$), while all three cases require comparable CNOT numbers in the trivial phase region. The nontrivial phase requires up to four times more CNOT gates than the trivial phase region.
    (d) Circuit depth, measured in number of layers, as a function of $\delta$ for the three different $T$. The maximum depths are 51 layers for $T = 20$, 106 layers for $T = 100$, and 89 layers for $T = 200$.
    Note that for \(\delta=\pm 1\), the model reduces to isolated dimers and dangling edges (for $\delta=-1$), leading to much simpler circuits.}
    \label{fig4}
\end{figure*}

Figure~\ref{fig3} demonstrates the robustness of the AVQDS algorithm in calculating the Berry phase by systematically analyzing its dependence on various parameters. Figure~\figref{fig3}{(a)} shows the computed Berry phase $\varphi_\mathrm{B}$ as a function of the cyclic state evolution time $T$ for dimerization parameters $\delta = -0.9$ (black circles) and $\delta = -0.5$ (orange diamonds). The ED result for $T=200$ is shown as a red dashed line. The Berry phase is accurately obtained for evolution times as short as $T = 1.6$ ($T = 2.0$) for $\delta = -0.9$ ($\delta = -0.5$). However, as shown in Fig.~\figref{fig3}{(d)}, the corresponding maximum infidelity during the evolution is large in this regime, with $\max_t[1 - f(t)]$ approaching 1. This highlights the robustness of the algorithm: despite strong nonadiabaticity, the Berry phase is computed accurately. The slightly lower critical $T$ required for $\delta = -0.9$ compared to $\delta = -0.5$ can be attributed to the larger energy gap $\Delta E_1$ between the ground state and first excited state for $\delta = -0.9$. Specifically, numerically we find that the energy gap between the ground state and the first excited state remains constant with varying $\lambda$, with $\Delta E_1 = 3.6$ for $\delta = -0.9$ and $\Delta E_1 = 2.0$ for $\delta = -0.5$ (see inset of Fig.~\figref{fig4}{(b)}). Since a larger gap allows for a shorter evolution time $T$ to keep the system close to the instantaneous ground state throughout the cyclic evolution, the critical $T$ is smaller for $\delta=-0.9$ than $\delta=-0.5$. Figure~\figref{fig3}{(g)} presents the corresponding number of CNOT gates. The minimum $T$ required for an accurate Berry phase calculation corresponds to a CNOT count of 264 for $\delta = -0.9$ and 244 for $\delta = -0.5$. The number of CNOT gates remains relatively stable as $T$ increases, showing only a minor overall decrease.  

Next, we examine the dependence of the Berry phase calculation on the time step size $\delta t$ used in solving the equation of motion for the variational parameters, Eq.~\eqref{eq:theta_eom}. Previous simulations employed an adaptive step size, whereas we now investigate the impact of a fixed $\delta t$ on the accuracy of the calculation. Figure~\figref{fig3}{(b)} shows the computed Berry phase as a function of $\delta t$ for a fixed $T = 20$. Results are presented for $\delta = -0.9$ (black circles) and $\delta = -0.5$ (orange diamonds) and are compared with the ED result for $T=200$ (red dashed line). A step size of $\delta t = 0.5$ is sufficient to obtain an accurate Berry phase. However, for $\delta t > 0.5$, the computed Berry phase significantly deviates from the expected value $\varphi_\mathrm{B} = \pi$. This trend is further supported by the corresponding maximum infidelity during the cyclic evolution shown in Fig.~\figref{fig3}{(e)}. Specifically, $\max_t[1 - f(t)]$ remains below 0.12 for $\delta t \leq 0.5$ but increases substantially for $\delta t > 0.5$, implying that the error in numerical integration dominates.  
The corresponding number of CNOT gates is shown in Fig.~\figref{fig3}{(h)}. For $\delta = -0.9$, the CNOT count starts at 416 for $\delta t = 0.01$ and gradually decreases until $\delta t \approx 0.35$. Beyond this threshold, where the accuracy of the Berry phase calculation deteriorates, the CNOT count turns to increase. In contrast, for $\delta = -0.5$, the CNOT count remains relatively stable up to the critical $\delta t$, beyond which it also increases. 

We further investigate the dependence of the Berry phase calculation on the McLachlan distance cutoff $L^2_\text{cut}$. Figure~\figref{fig3}{(c)} presents the computed Berry phase as a function of $L^2_\text{cut}$ for a fixed evolution time $T = 20$. Results are shown for $\delta = -0.9$ (black circles) and $\delta = -0.5$ (orange diamonds), alongside the ED results (red dashed line). The Berry phase remains accurate for $L^2_\text{cut} \leq 10^{-1}$ ($L^2_\text{cut} \leq 10^{-2}$) for $\delta=-0.9$ ($\delta=-0.5$), but strong deviations emerge beyond these thresholds. This trend is further confirmed by the maximum infidelity shown in Fig.~\figref{fig3}{(f)}. The quantity $\max_t[1 - f(t)]$ remains low below the critical $L^2_\text{cut}$ but increases significantly above it, reflecting the limited expressibility of the automatically generated ansatz. The corresponding number of CNOT gates is plotted in Fig.~\figref{fig3}{(i)}. For both values of $\delta$, the number of CNOT gates decreases as $L^2_\text{cut}$ increases. Notably, when $L^2_\text{cut}$ increases from $10^{-6}$ to its critical value, the CNOT count is reduced by approximately a factor of 4 (3) for $\delta=-0.9$ ($\delta=-0.5$), significantly lowering computational cost.  Overall, the dependencies on $T$, $\delta t$, and $L^2_\text{cut}$ highlight the robustness of the AVQDS algorithm: even when $T$ is decreased significantly and $\delta t$ and $L^2_\text{cut}$ are increased substantially, resulting in strong nonadiabaticity or higher infidelity in the simulations, the Berry phase remains accurately determined.

\begin{figure*}[t!]
    \centering
    \includegraphics[scale=0.38]{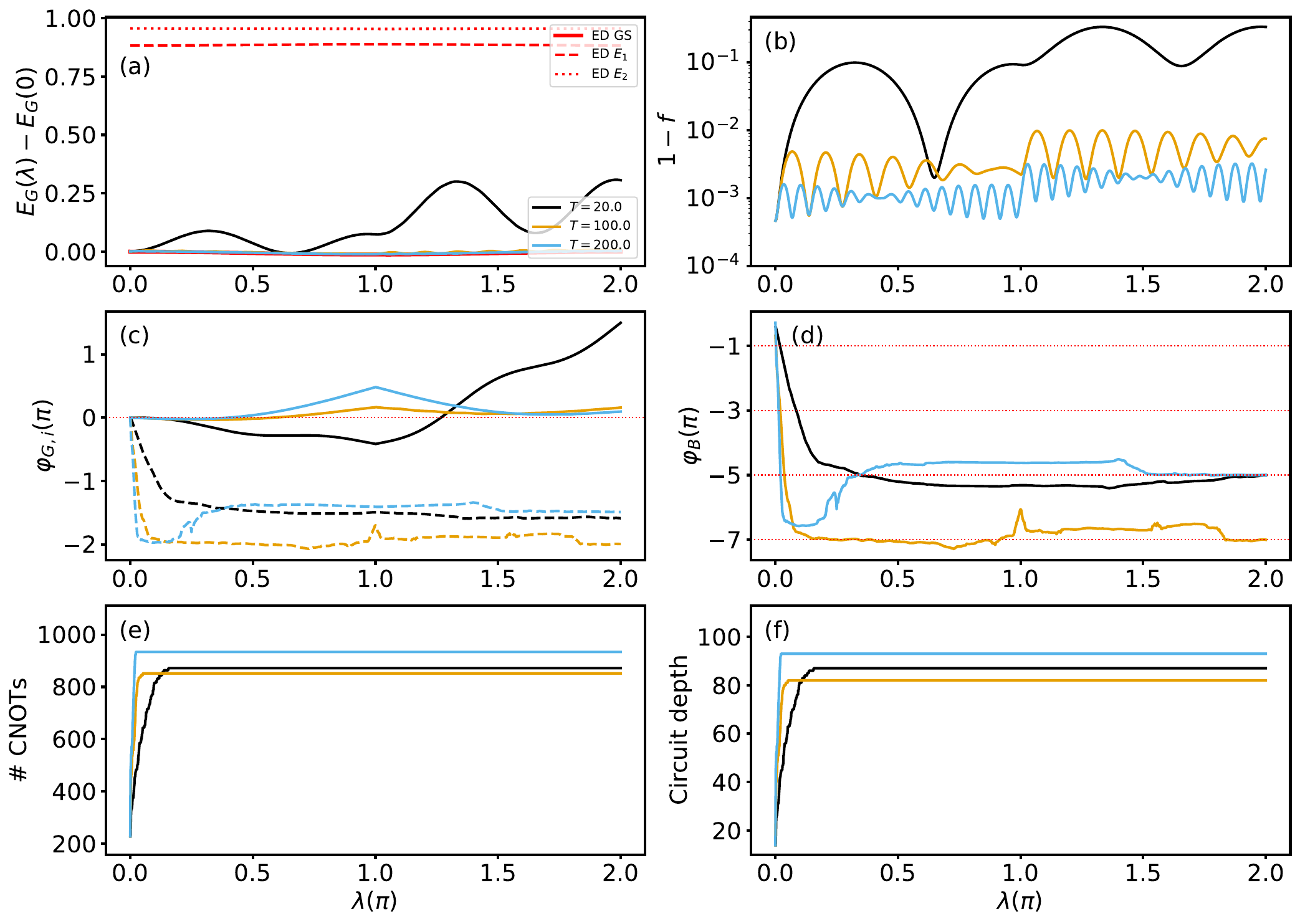}
    \caption{Benchmarking of the AVQDS approach for calculating the Berry phase of the interacting SSHH model with $U=10$. 
    (a) The quantum state energy $E(\lambda)$ for cyclic state evolution times of $T=20$ (black line), $T=100$ (orange line), and $T=200$ (blue line), compared to exact diagonalization results (red line). Results are shown for a dimerization parameter of $\delta=-0.6$. For comparison,  the first two ED excited energies $E_{1,2}(\lambda)$ are plotted in red with distinct dashed line styles. 
    (b) Corresponding infidelity $1-f$ as a function of $\lambda$, which decreases with increasing $T$. 
    (c) Contributions to the global phase $\varphi_G$: dynamical phase $\varphi_{\mathrm{G},1}(\lambda)$ (solid lines) and geometric phase $\varphi_{\mathrm{G},2}(\lambda)$ (multiplied by $0.1$, dashed lines). The red dashed line marks $\varphi_{\mathrm{G},i} = 0$. While $\varphi_\mathrm{G,1}$ approaches zero for large $T$, $\varphi_\mathrm{G,2}$ exhibits residual deviations even for $T=200$. 
    (d) Berry phase calculated using Eq.~\eqref{eq:phiB2} for the three different values of $T$. The red dashed lines indicate $\varphi_\mathrm{B}=-(2n+1)\pi$ with $n=0,1,2,3$. Note that bringing the angle to its principal value, $\varphi_\mathrm{B, principal}=\varphi_\mathrm{B} -2\pi \,\text{round}[\varphi_\mathrm{B}/(2\pi)]$ yields $\varphi_\mathrm{B, principal}=\pi$ for all three values of $T$. This result matches the ED value $\varphi_\mathrm{B}=\pi$, confirming the robustness of the AVQDS approach.
    (e) CNOT gate count as a function of $\lambda$ for the three different values of $T$, showing saturation beyond $\lambda=0.16\pi$. The CNOT counts are comparable across different $T$ and indicate increased complexity compared to the noninteracting case in Fig.~\figref{fig2}{(e)}. The maximum CNOT counts are 872, 852, and 934 for $T = 20$, $T=100$, and $T = 200$, respectively. 
    (f) Circuit depth in terms of the number of layers of unitaries. The depth reaches 87 layers for $T = 20$, 82 layers for $T = 100$, and 93 layers for $T = 200$.}
    \label{fig5}
\end{figure*}

To further assess the accuracy of the AVQDS approach in capturing topological properties, we compute the Berry phase $\varphi_\mathrm{B}$ as a function of the dimerization parameter $\delta$ and compare it to ED results. Figure~\figref{fig4}{(a)} presents the Berry phase obtained using the AVQDS approach for different total evolution times $T$--specifically, $T = 20$ (black circles), $T = 100$ (orange diamonds), and $T = 200$ (blue triangles)--alongside ED results (gray squares). All AVQDS simulations accurately reproduce the ED result and successfully resolve the phase transition at $\delta = 0$.  
Figure~\figref{fig4}{(b)} shows the corresponding maximum infidelity during the cyclic state evolution as a function of $\delta$ for the three different values of $T$. The infidelity increases as $\delta$ approaches the phase transition at $\delta = 0$ from both negative and positive $\delta$ values, highlighting the challenge of maintaining adiabaticity near the critical point. As shown in the inset of Fig.~\figref{fig4}{(b)}, the low-energy spectrum under the twisted boundary $\lambda=\pi$ exhibits a decreasing gap $\Delta E_1$ as $\delta\!\to\!0$, which accounts for the infidelity peak and the increased resources required to remain adiabatic near the transition at $\delta=0$. Additionally, the maximum infidelity $\max_t[1-f(t)]$ is higher in the nontrivial Berry phase where $\varphi_\mathrm{B}=\pi$. Extending the evolution time reduces the maximum infidelity, from approximately 0.48 for $T = 20$ to 0.04 for $T = 200$, confirming the convergence toward adiabatic state evolution.

\begin{figure*}[t!]
    \centering
    \includegraphics[width=\textwidth]{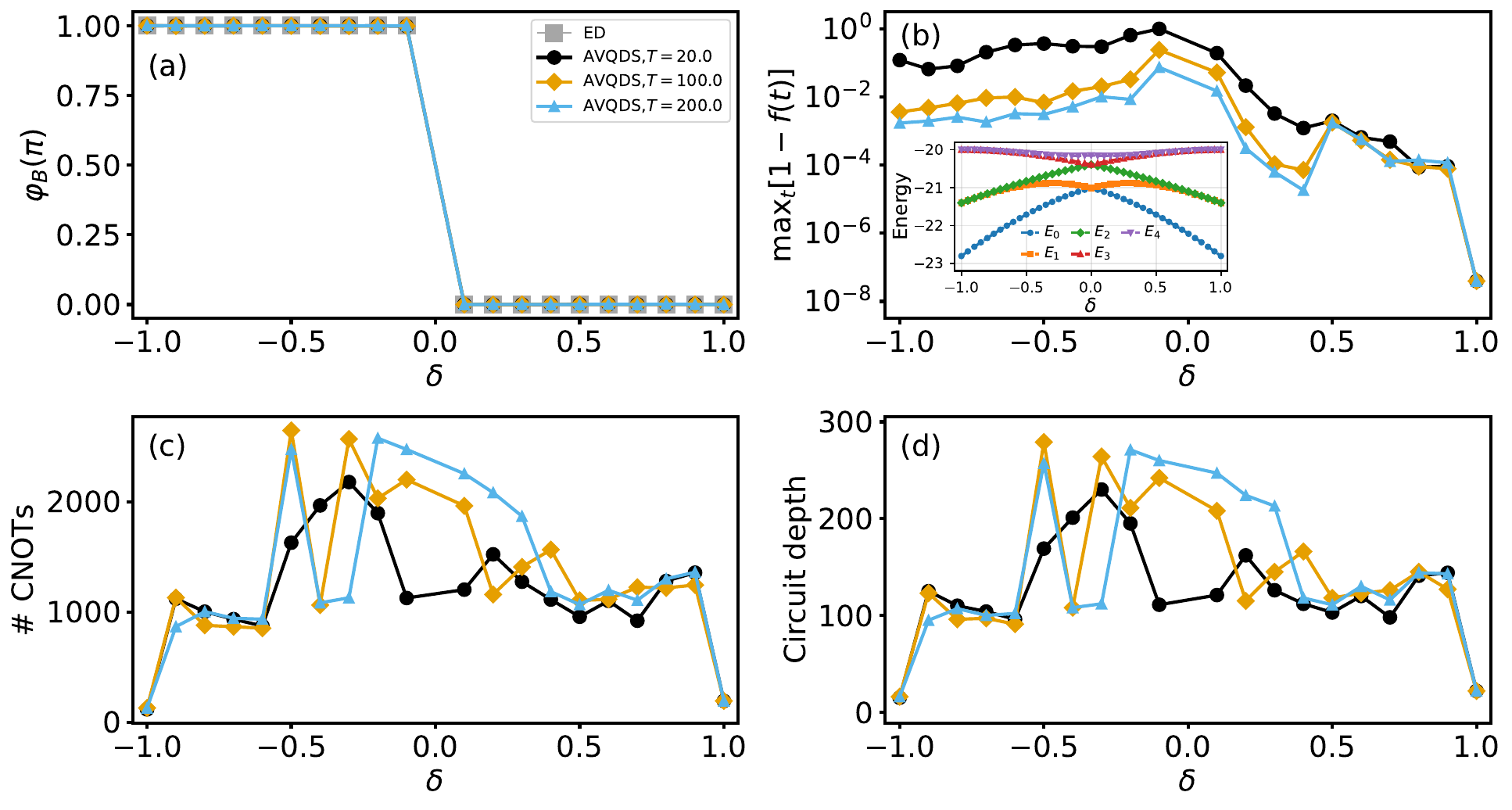}
    \caption{Quantum circuit complexity and fidelity in the Berry phase calculation across a topological phase transition for the interacting SSHH model. 
    (a) Berry phase $\varphi_\mathrm{B}$ as a function of the dimerization parameter $\delta$, computed using the AVQDS approach for evolution times $T=20$ (black circles), $T=100$ (orange diamonds), and $T=200$ (blue triangles), compared to exact diagonalization results (ED, gray squares). 
    (b) The corresponding maximum infidelity during adiabatic evolution, $\max_t[1-f(t)]$, demonstrating increased infidelity near the phase transition at $\delta = 0$ and larger $\max_t[1-f(t)]$ in the non-trivial Berry-phase region ($\varphi_\mathrm{B}=\pi$) compared to the trivial phase region ($\varphi_\mathrm{B}=0$), consistent with the noninteracting SSH model results in Fig.~\figref{fig4}{(b)}. Inset: Lowest five many-body energies $E_{0\ldots4}$ vs. dimerization $\delta$ under a twisted boundary with $\lambda=\pi$. The gap narrowing as $\delta \to 0$ motivates larger $T$ to mitigate nonadiabaticity near the topological phase transition at $\delta=0$.
    (c, d) Corresponding CNOT gate count and circuit depth as a function of $\delta$, illustrating the increased complexity of the interacting model compared to the noninteracting case.
    Observe that at $\delta = \pm 1$, the system simplifies to isolated dimers with dangling edge sites (for $\delta = -1$), resulting in simplest circuits.
    }
    \label{fig6}
\end{figure*}

Figure~\figref{fig4}{(c)} shows the number of required CNOT gates as a function of $\delta$. Among the three evolution times, $T = 20$ requires the fewest CNOT gates, while $T = 100$ results in the highest CNOT count across most of the non-trivial topological region ($\varphi_\mathrm{B} = \pi$). In contrast, in the trivial Berry phase region ($\varphi_\mathrm{B} = 0$), the number of CNOT gates remains comparable across all three values of $T$. Notably, the non-trivial Berry phase region requires significantly more CNOT operations--up to four times as many--compared to the trivial phase region. Figure~\figref{fig4}{(d)} presents the corresponding circuit depth, measured by the number of layers of unitaries, as a function of $\delta$. The circuit depth reaches a maximum of 51 layers for $T = 20$, while for $T = 100$ and $T = 200$ it increases to 106 and 89 layers, respectively.

\subsubsection{Interacting SSHH Model}

We now analyze the performance and quantum resource requirements of the algorithm described in Sec.~\ref{sec:methods} for the interacting SSHH model with $U=10$ in the strongly correlated regime. Analogous to the noninteracting case, we employ the Hamiltonian operator pool in the AVQDS calculations, consisting of $N_\mathrm{p}=22$ operators for a system of $N=4$ sites. Figure~\figref{fig5}{(a)} presents the energy $E_\mathrm{G}(\lambda)$ of the state propagated according to AVQDS for $T=20$ (black line), $T=100$ (orange line), and $T=200$ (blue line), compared to the ground state energy calculated with ED (red line). The results are presented for a dimerization parameter of $\delta=-0.6$. In agreement with the observations for the noninteracting model (Fig.~\figref{fig2}{(a)}), $E_\mathrm{G}(\lambda)$ converges toward the exact result as $T$ increases. However, deviations from the exact energy are more pronounced in the interacting case, which can be attributed to the reduced energy gap caused by correlations, making adiabatic evolution more challenging. This complexity is further confirmed by the corresponding infidelity plotted in Fig.~\figref{fig5}{(b)}. For $T=20$, the infidelity is approximately six times larger than in the noninteracting model (Fig.~\ref{fig2}{(b)}), peaking around $0.33$. As $T$ increases, the infidelity decreases, with a maximum value of only $3.2\times 10^{-3}$ for $T=200$.

The same interference mechanism discussed for the noninteracting case explains the infidelity oscillations in Fig.~\figref{fig5}{(b)}. With a nearly constant lowest gap $\Delta E_1\simeq 0.9$ over the loop (Fig.~\figref{fig5}{(a)}) and a linear ramp $\lambda(t)=2\pi t/T$, the $\lambda$-domain frequency is $\Omega_\lambda=\Delta E_1 T/(2\pi)$, giving $\Omega_\lambda\approx 2.9,\,14.3,\,28.6$ and periods $\Delta\lambda/\pi=2/\Omega_\lambda\approx 0.70,\,0.14,\,0.07$ for $T=20,100,200$, in agreement with Fig.~\figref{fig5}{(b)}.

The contributions to the Berry phase $\varphi_\mathrm{B}$ from the global phase $\varphi_\mathrm{G}$ are analyzed in Fig.~\figref{fig5}{(c)}, where $\varphi_\mathrm{G,1}(\lambda)$ is shown as solid lines and $\varphi_\mathrm{G,2}(\lambda)$ (scaled by $0.1$) is plotted as dashed lines for the three different values of $T$. As $T$ increases, $\varphi_\mathrm{G,1}$ approaches zero, indicating that the dynamical phase largely cancels out during the cyclic state evolution. However, unlike in the noninteracting case (Fig.~\figref{fig2}{(c)}), $\varphi_\mathrm{G,2}$ still exhibits deviations from zero even for $T=200$. Figure~\figref{fig5}{(d)} presents the Berry phase $\varphi_\mathrm{B}$ calculated using Eq.~\eqref{eq:phiB2}, with the phase $\varphi_\mathrm{qc}$ obtained from the quantum circuit in Fig.~\figref{fig1}{(b)} and $\varphi_\mathrm{G}$ computed via Eq.~\eqref{eq:phiG}. The dashed red lines indicate $\varphi_\mathrm{B}=-(2n+1)\pi$ for $n=0,1,2,3$. The calculated $\varphi_\mathrm{B}$ 
exceeds its principal value range $(-\pi, \pi]$. Nevertheless, the obtained Berry phase is $\varphi_\mathrm{B}=-(2n+1)\pi$ with $n=1$ for $T=20$ and $T=200$ while $n=2$ for $T=100$. Applying the principal value transformation, $\varphi_\mathrm{B, principal}=\varphi_\mathrm{B} -2\pi \,\text{round}[\varphi_\mathrm{B}/(2\pi)]$ results in $\varphi_\mathrm{B, principal}=\pi$ for all three values of $T$. This agrees with the ED result $\varphi_\mathrm{B}=\pi$. As a result, despite non-vanishing $\varphi_\mathrm{G}$ across different $T$ (Fig.~\figref{fig5}{(c)}), the Berry phase results in Fig.~\figref{fig5}{(d)} remain accurate, confirming the robustness of the employed approach.

Figure~\figref{fig5}{(e)} shows the required number of CNOT gates as a function of $\lambda$ for the three different values of $T$. Ground-state preparation via AVQITE requires approximately 92 CNOT gates, yielding an infidelity of $4.7\times 10^{-4}$. The CNOT count increases with $\lambda$, primarily during the initial evolution stage, before saturating near $\lambda=0.16\pi$. The saturated CNOT count remains comparable across the three studied values of $T$. Specifically, for $T=20$, the CNOT count reaches 872 at $\lambda=2\pi$, while for $T=100$ and $T=200$, the maximum CNOT count is 852 and 934, respectively. The corresponding circuit depth is shown in Fig.~\figref{fig5}{(f)}. Throughout the cyclic state evolution, the circuit depth grows from an initial 9 layers generated by AVQITE to a maximum of 87 layers for $T=20$, while for $T=100$ and $T=200$ the depths reach 82 and 93 layers, respectively.

Finally, we analyze the Berry phase as a function of the dimerization parameter $\delta$ to investigate the presence of a topological phase transition and further benchmark the performance of the AVQDS approach, analogous to the noninteracting case studied in Fig.~\ref{fig4}.
Figure~\figref{fig6}{(a)} presents the Berry phase $\varphi_\mathrm{B}$ as a function of $\delta$. The results obtained using AVQDS with $T=20$ (black circles), $T=100$ (yellow diamonds), and $T=200$ (blue triangles) are compared to those from ED (gray squares). All AVQDS results agree with the ED values of $\varphi_\mathrm{B}$ and successfully capture the phase transition occurring at $\delta = 0$.
The corresponding maximum infidelity during the cyclic state evolution, $\max_t[1-f(t)]$, is shown in Fig.~\figref{fig6}{(b)}. In agreement with the results for the noninteracting case in Fig.~\figref{fig4}{(b)}, the infidelity increases as $\delta$ approaches the phase transition at $\delta = 0$. The maximum infidelity decreases from approximately 1.0 for $T=20$ to 0.074 for $T=200$, demonstrating improved accuracy with increasing $T$. This behavior is consistent with the energy spectrum in the inset of Fig.~\figref{fig6}{(b)}: under the twisted boundary $\lambda=\pi$, the low-energy gap shrinks as $\delta\to 0$, requiring larger $T$ to maintain comparable fidelity (due to enhanced nonadiabaticity) and thereby increasing resource usage, as discussed next.
Figure~\figref{fig6}{(c)} shows the dependence of the CNOT gate count on $\delta$. Similar to the noninteracting case, the $\delta$-region associated with a nontrivial Berry phase can require a higher number of CNOT gates than the region with a trivial Berry phase. However, no clear trend is observed with respect to the evolution time $T$. The maximum CNOT count reaches 2178 for $T=20$ and 2646 (2576) for $T=100$ ($T=200$), highlighting the increased quantum resource requirements of the interacting model compared to the noninteracting model in Fig.~\figref{fig4}{(c)}.
The circuit depth as a function of $\delta$ is plotted in Fig.~\figref{fig6}{(d)}. The circuit depth reaches up to 230 layers for $T=20$, 279 layers for $T=100$, and 271 layers for $T=200$, which is approximately three times deeper than the circuits required for the noninteracting SSH model.

\section{\label{sec:con} Conclusion and outlook}
In this work, we presented and benchmarked a quantum computing approach for calculating the Berry phase in topological Hamiltonian systems via cyclic state evolution. Our method leverages adaptive variational quantum algorithms for both ground state preparation and real-time propagation, optimizing circuit depth and gate complexity. Specifically, following the approach in Ref.~\citenum{Murta2020}, we constructed the cyclic adiabatic evolution operator $U_\text{loop}$, replacing conventional Trotter-based evolution with AVQDS for circuit compression. Furthermore, we utilized AVQITE to efficiently prepare the ground state at $\lambda = 0$, ensuring accurate initialization for the Berry phase calculation. This combination significantly reduces circuit depth while maintaining high accuracy~\cite{AVQDS, AVQITE, mootz2024adaptive}.

To benchmark AVQDS for Berry phase computations, we studied a four-site dimerized Fermi-Hubbard chain and validated our results with exact diagonalization. Our findings demonstrate that AVQDS accurately captures the Berry phase in both noninteracting and interacting regimes, successfully resolving topological phase transitions. Notably, the method remains reliable even when the wavefunction fidelity is relatively low, underscoring its robustness in evaluating the nonlocal topological properties. The circuit complexity depends on electron correlation strength: up to 92 layers in the noninteracting case and up to 120 layers in the interacting model.

We further demonstrated the robustness of our approach by examining its dependence on cyclic simulation time $T$, time step $\delta t$, and McLachlan distance cutoff $L^2_\text{cut}$. Our results indicate that accurate Berry phase calculations are achievable with moderate values of $T \geq 2$, $\delta t \leq 0.5$, and $L^2_\text{cut} \leq 10^{-2}$ for the noninteracting SSH model. The minimal $T$ required is directly linked to the energy gap between the ground and first excited states, as adiabatic evolution necessitates a quench rate $\propto 1/T$ slower than this gap.

Two well-studied families of accelerated adiabatic protocols are compatible with our AVQDS framework and could reduce the required number of integration steps at a fixed accuracy, with model- and accuracy-dependent trade-offs. Counterdiabatic (adiabatic-gauge-potential) driving augments the target Hamiltonian by a term $\h_{\mathrm{CD}}(\lambda,\dot\lambda)$ that suppresses diabatic transitions and tracks the adiabatic trajectory even for finite $T$~\cite{Berry2009,Sels2017,Claeys2019,Odelin2019}. In our setting this corresponds to evolving with $\h(\lambda)\!\to\!\h(\lambda)+\h_{\mathrm{CD}}(\lambda,\dot\lambda)$ within AVQDS: while the total time $T$ (and thus the RK4 step count) can decrease, the added generators in $\h_{\mathrm{CD}}$ generally increase the per-step measurement/gate cost, so the net depth change is model- and accuracy-dependent. Composite/$\pi$-pulse shortcuts provide an alternative acceleration by inserting brief unitary pulses at prescribed times to suppress diabatic leakage and control errors while preserving the same closed geometric path in $\lambda$~\cite{Torosov2011,Balasubramanian2018,Odelin2019}. Within AVQDS, this is handled by (i) integrating the parameter equations of motion on each interval between pulses, (ii) applying the pulse unitary to update the state/parameters at each pulse time, and (iii) continuing the integration on the next interval. Such sequences can reduce $T$ with only a small constant overhead from added single-qubit rotations, leaving two-qubit counts largely unchanged unless entangling operations are explicitly included. A systematic benchmark of these accelerated schedules on the SSHH problems studied here is a natural direction for future work.

Our findings highlight the potential of variational quantum algorithms for tackling complex quantum geometric phase problems. Future work includes extending AVQDS to larger system sizes and more intricate topological models, such as interacting topological insulators and superconductors. This necessarily hinges on the ongoing efforts toward enabling AVQDS calculations on quantum hardware, including error mitigation techniques~\cite{caiQuantumErrorMitigation2022, Getelina2024QuantumSE, chen2025quantum}, strategies for optimal shot allocation~\cite{van2021measurement}, and the synergistic integration of classical and quantum resources~\cite{Zhang2025TETRIS, chen2024minimally}. Extending this approach to quantum embedding frameworks~\cite{bauer2020quantum, gqce, chen2025quantum} is a promising direction, given the substantially reduced quantum resource requirements for simulating bulk systems. The capability to efficiently evaluate geometric phases using variational quantum algorithms may open pathways to studying a wide range of correlated topological quantum phenomena that are intractable for classical methods.

\section*{Acknowledgment}
We are grateful for discussions with A. Khindanov, T. Iadecola, F. Zhang, and C.-Z. Wang. 
This work was supported by the U.S. Department of Energy (DOE), Office of Science, Basic Energy Sciences, Materials Science and Engineering Division, including the grant of computer time at the National Energy Research Scientific Computing Center (NERSC) in Berkeley, California. The research was performed at the Ames National Laboratory, which is operated for the U.S. DOE by Iowa State University under Contract No. DE-AC02-07CH11358.

\section*{Data Availability Statement}
The data that support the findings of this study are openly available at figshare~\cite{Mootz2025data_berry}.

\appendix

\section{Global phase contribution in the adiabatic limit \label{sec:app}}

In this appendix, we prove that the global phase contributions $\varphi_\mathrm{G,i}$ defined in Eq.~\eqref{eq:phiGi} individually vanish in numerically exact cyclic adiabatic quantum dynamics simulations ($T \to \infty$) and for time-reversal symmetric variational parameters. This holds under the assumption that the system follows the instantaneous ground state and that the adaptive expansion of the variational ansatz during the dynamics does not introduce asymmetries between both half-cycles of the cyclic adiabatic evolution. The first contribution, $\varphi_\mathrm{G,1}$ corresponds to the dynamical phase and is determined by the ground state energy in the variational state $E_{\mathrm{G},\bth}\equiv \av{\h}_\theta$. In the first half cycle, the dynamical phase is given by
\begin{align}
    \varphi_\mathrm{G,1}^{+} &= \int_{0}^{T/2}\mathrm{d}t\, E_{\mathrm{G},\bth}(t)=\int_{\lambda(0)}^{\lambda(T/2)}\mathrm{d}\lambda\,E_{\mathrm{G},\bth}(\lambda)\frac{\mathrm{d}t}{\mathrm{d}\lambda}\nonumber \\ 
    &= \frac{T}{2\pi}\int_0^\pi \mathrm{d}\lambda\, E_{\mathrm{G},\bth}(\lambda)\,,
\end{align}
where we used $\lambda(t)=\frac{2\pi}{T} t$ in the last step. In the second half cycle, the dynamical phase is given by
\begin{align}
    \varphi_\mathrm{G,1}^{-} &= \int^{0}_{T/2}\mathrm{d}t\,E_{\mathrm{G},\bth}(t) = \int^{\lambda(0)}_{\lambda(T/2)}\mathrm{d}\lambda\,E_{\mathrm{G},\bth}(\lambda)\frac{\mathrm{d}t}{\mathrm{d}\lambda}\nonumber \\ 
    &= -\frac{T}{2\pi}\int_\pi^{2\pi} \mathrm{d}\lambda\, E_{\mathrm{G},\bth}(\lambda)= -\frac{T}{2\pi}\int_{-\pi}^{0} \mathrm{d}\lambda\, E_{\mathrm{G},\bth}(\lambda) \nonumber \\ 
    &= -\frac{T}{2\pi}\int_0^{\pi} \mathrm{d}\lambda\, E_{\mathrm{G},\bth}(-\lambda)\,,
\end{align}
where we used $\lambda(t)=\frac{2\pi}{T}(T - t)$ in the second step. As a result, the dynamical phase contribution to the global phase, $\varphi_\mathrm{G,1}=\varphi_\mathrm{G,1}^{+}+\varphi_\mathrm{G,1}^{-}$ vanishes for $E_{\mathrm{G},\bth}(\lambda) = E_{\mathrm{G},\bth}(-\lambda)$ which holds for Hamiltonians with time-reversal symmetry. In the simulations, this condition is realized for sufficient large evolution time $T$, ensuring the system follows the instantaneous ground state, as demonstrated in Figs.~\ref{fig2} and \ref{fig5}. 

The second contribution to the global phase, $\varphi_\mathrm{G,2}$, is determined by the geometric properties of the cyclic adiabatic evolution. In the first half cycle, this contribution is
\begin{align}
    \varphi_\mathrm{G,2}^{+}&=-i\int_0^{T/2}\mathrm{d}t\, \mathrm{Im} \left[ \sum_\mu\bra{\Psi[\bth]}\frac{\partial\ket{\Psi[\bth]}}{\partial \theta_\mu}\frac{\mathrm{d}\theta_\mu}{\mathrm{d}t} \right]\nonumber \\
    &=-i\int_0^{\pi}\mathrm{d}\lambda\, \mathrm{Im} \left[\bra{\Psi[\bth(\lambda)]}\frac{\partial\ket{\Psi[\bth(\lambda)]}}{\partial \lambda}\right]\,.
\end{align}
For the second half of the adiabatic cycle we find
\begin{align}
    \varphi_\mathrm{G,2}^{-}&=-i\int^0_{T/2}\mathrm{d}t\, \mathrm{Im} \left[\sum_\mu\bra{\Psi[\bth]}\frac{\partial\ket{\Psi[\bth]}}{\partial \theta_\mu}\frac{\mathrm{d}\theta_\mu}{\mathrm{d}t} \right]\nonumber \\
    &= -i\int_\pi^{2\pi}\mathrm{d}\lambda\, \mathrm{Im} \left[\bra{\Psi[\bth(\lambda)]}\frac{\partial\ket{\Psi[\bth(\lambda)]}}{\partial \lambda} \right]\nonumber \\
    &= i\int_0^{\pi}\mathrm{d}\lambda\,\mathrm{Im} \left[\bra{\Psi[\bth(-\lambda)]}\frac{\partial\ket{\Psi[\bth(-\lambda)]}}{\partial \lambda}\right]\,.
\end{align}
As a result, the second contribution to the global phase, $\varphi_\mathrm{G,2}=\varphi_\mathrm{G,2}^{+}+\varphi_\mathrm{G,2}^{-}$, vanishes  
for variational parameters satisfying time-reversal symmetry such that $\ket{\Psi[\bth(\lambda)]}=\ket{\Psi[\bth(-\lambda)]}$. This condition holds for Hamiltonians with time-reversal symmetry and a fixed variational ansatz. In AVQDS, where the variational ansatz dynamically expands during adiabatic evolution, this is approximately realized for sufficient large evolution time $T$ with unitaries primarily appended to the variational ansatz during the first half of the adiabatic evolution, as observable in Fig.~\ref{fig2}.  In this situation, the ansatz is effectively fixed  in the second half of the adiabatic cycle. Consequently, for sufficiently large $T$, the evolution in the second half becomes a mirrored evolution (under time-reversal symmetry) of the first half. Thus, the contributions to the geometric phase from the second half of the cycle cancel those from the first half, leading to $\varphi_\mathrm{G,2} = 0$. In practice, $\varphi_\mathrm{G,2}$ is often nonzero due to deviations from perfect time-reversal symmetry of the variational parameters, caused by finite $T$ and numerical errors. However, we find that this finite contribution is compensated by the quantum-circuit-based phase $\varphi_\mathrm{qc}$, calculated using the circuit in Fig.~\figref{fig1}{(b)}. As a result, the total phase $\varphi_\mathrm{qc}+\varphi_\mathrm{G}$ accurately yields the Berry phase $\varphi_\mathrm{B}$, ensuring the reliability of the quantum simulation approach to retrieve the topological properties.

\bibliography{ref}

\end{document}